\documentclass[linenumbers]{aastex631}

\usepackage{chngcntr}
\usepackage{soul}
\usepackage{bm}

\begin{document}

\title{HDO ice detected toward an isolated low-mass protostar with JWST}

\author[0000-0002-7433-1035]{Katerina Slavicinska}
\affiliation{Laboratory for Astrophysics, Leiden Observatory, P.O. Box 9513, 2300 RA Leiden, NL}

\author[0000-0002-9470-2358]{\L{}ukasz Tychoniec}
\affiliation{Leiden Observatory, P.O. Box 9513, 2300 RA Leiden, NL}

\author[0000-0002-1860-2304]{Mar\'ia Gabriela Navarro}
\affiliation{INAF -- Osservatorio Astronomico di Roma, Via di Frascati 33, 00040, Monteporzio Catone, IT}

\author[0000-0001-7591-1907]{Ewine F. van Dishoeck}
\affiliation{Leiden Observatory, P.O. Box 9513, 2300 RA Leiden, NL}
\affiliation{Max Planck Institut f\"ur Extraterrestrische Physik (MPE), Giessenbachstrasse 1, 85748 Garching, DE}

\author[0000-0002-6195-0152]{John J. Tobin}
\affiliation{National Radio Astronomy Observatory, 520 Edgemont Rd., Charlottesville, VA 22903, USA}

\author[0000-0002-6312-8525]{Martijn L. van Gelder}
\affiliation{Leiden Observatory, P.O. Box 9513, 2300 RA Leiden, NL}

\author[0000-0002-3395-5634]{Yuan Chen}
\affiliation{Leiden Observatory, P.O. Box 9513, 2300 RA Leiden, NL}

\author[0000-0001-9344-0096]{A. C. Adwin Boogert}
\affiliation{Institute for Astronomy, University of Hawai’i at Manoa, 2680 Woodlawn Drive, Honolulu, HI 96822, USA}

\author[0009-0008-2128-6040]{W. Blake Drechsler}
\affiliation{Department of Astronomy, University of Virginia, 530 McCormick Rd., Charlottesville, VA, 22904, USA}

\author[0000-0002-1700-090X]{Henrik Beuther}
\affiliation{Max Planck Institute for Astronomy, Heidelberg, Baden-W\"urttemberg, DE}

\author[0000-0001-8876-6614]{Alessio Caratti o Garatti}
\affiliation{INAF -- Osservatorio Astronomico di Capodimonte, Salita Moiariello 16, 80131 Napoli, IT}

\author[0000-0001-7629-3573]{S. Thomas Megeath}
\affiliation{Ritter Astrophysical Research Center, Department of Physics and Astronomy, University of Toledo, Toledo, OH 43606, USA}

\author[0000-0001-9443-0463]{Pamela Klaassen}
\affiliation{UK Astronomy Technology Centre, Royal Observatory Edinburgh, Blackford Hill, Edinburgh EH9 3HJ, UK}

\author[0000-0002-4540-6587]{Leslie W. Looney}
\affiliation{National Radio Astronomy Observatory, 520 Edgemont Rd., Charlottesville, VA 22903, USA}
\affiliation{Department of Astronomy, University of Illinois, 1002 West Green St, Urbana, IL 61801, USA}

\author[0000-0001-6872-2358]{Patrick J. Kavanagh}
\affiliation{Department of Physics, Maynooth University, Maynooth, Co. Kildare, Ireland}

\author[0000-0001-7826-7934]{Nashanty G. C. Brunken}
\affiliation{Leiden Observatory, P.O. Box 9513, 2300 RA Leiden, NL}

\author[0000-0002-9209-8708]{Patrick Sheehan}
\affiliation{National Radio Astronomy Observatory, 520 Edgemont Rd., Charlottesville, VA 22903, USA}

\author[0000-0002-3747-2496]{William J. Fischer}
\affiliation{Space Telescope Science Institute, 3700 San Martin Dr, Baltimore, MD 21218, USA}



\begin{abstract}

Water is detected in environments representing every stage of star and solar system formation, but its chemical evolution throughout these stages remains poorly constrained. Deuterium ratios offer a means of probing chemical links between water in different cosmic regions because of their sensitivity to physicochemical conditions. Here, we present the first detection of the 4.1 $\mu$m HDO ice feature with JWST toward a low-mass protostar, L1527 IRS, which may eventually grow to a sun-like mass. We measure an ice HDO/H$_{2}$O ratio of 4.4$^{+3.7}_{-1.7}$$\times$10$^{-3}$, where the reported error is dominated by uncertainties in continuum definition and ice band strengths. This fraction is similar to the gas HDO/H$_{2}$O ratios measured in the warm ($>$100 K) inner cores of other low-mass protostellar envelopes and protoplanetary disks found in comparably isolated star-forming regions. Such a similarity tentatively supports the assumption that water vapor detected in these regions is not significantly altered by gas-phase reactions following ice sublimation. It also supports the hypothesis that pre- and protostellar water ice is largely inherited in a chemically unaltered state by outer protoplanetary disks. However, the fraction is a factor of $\sim$4-10 times higher than the gas HDO/H$_{2}$O ratios measured toward comets and low-mass protostars in clustered star-forming regions. This difference may be due to either gas-phase water reprocessing in protostellar envelopes and protoplanetary disks, or differences between prestellar conditions of isolated dense cores and the clustered star-forming regions that are more analogous to the environment in which our Sun formed.

\end{abstract}

\keywords{Astrochemistry (75) --- Protostars (1302) --- Isotopic abundances (867) --- Ice spectroscopy (2250) -- Infrared spectroscopy (2285)}


\section{Introduction} \label{sec:intro}

Water is a highly influential molecule that is essential to the formation and evolution of stars, planetary bodies, and life as we know it. It is also one of the most abundantly detected molecules in star-forming regions and has been observed at every evolutionary stage of star formation, from prestellar clouds to protostellar envelopes, protoplanetary disks, and solar system bodies such as our Earth, icy moons, and comets \citep{encrenaz2008water,van2014water,van2021water}.

In recent years, numerous works have investigated whether the water found in these various environments is chemically linked. Many of these studies examined water deuterium ratios because the sensitivity of deuterium isotopic exchange reactions to physicochemical conditions makes D/H ratios excellent tracers of the origins and evolutionary histories of molecules. In particular, much of the water detected in our solar system is D-enriched (e.g., \citealt{lecuyer1998hydrogen,waite2009liquid,alexander2012provenances,altwegg201567p}), in some cases by almost two orders of magnitude, with respect to the D/H observed in the diffuse interstellar medium ($\sim$2$\times$10$^{-5}$, \citealt{prodanovic2010deuterium}). Because deuterium enrichment is efficient in low-temperature environments, these high water D/H ratios may be a remnant signature of water's origins in the primitive ices of extremely cold and dense prestellar clouds from which stars and solar systems eventually form.

To confirm such a link between the water in our solar system and interstellar water ice, the water deuterium ratios (HDO/H$_{2}$O) of ices in young star-forming regions must be measured. In theory, this value can be obtained by calculating HDO ice column densities via the 4.1 $\mu$m O-D stretching mode of HDO. However, prior to the launch of the \textit{James Webb} Space Telescope (JWST), reported detections of this feature (e.g., \citealt{teixeira1999discovery}) remained controversial due to the low sensitivity of previous IR telescopes \citep{dartois2003revisiting,parise2003search}, hindering the direct quantification of reliable ice HDO/H$_{2}$O ratios.

Interferometric (sub)millimeter gas-phase studies thus attempted to quantify interstellar ice HDO/H$_{2}$O ratios indirectly instead by targeting the warm ($>$100 K) inner regions of protostellar envelopes (``hot corinos''). Because it is often argued that the majority of the gas species detected in these environments originate from the thermal desorption of ices with little subsequent gas-phase chemical alteration (e.g., \citealt{requena2006organic,coletta2020evolutionary,van2020complex,nazari2021complex,chen2023coccoa}), their measured gas HDO/H$_{2}$O ratios are assumed by some to reflect the ratios of their ice envelopes. Toward four protostars located in the clustered cloud complexes NGC 1333 and Ophiuchus, the measured gas HDO/H$_{2}$O ratios range between 5-9$\times$10$^{-4}$ \citep{persson2014deuterium}, very similar to the ratios measured in several comets in our solar system (e.g., \citealt{eberhardt1995d,villanueva2008sensitive,altwegg201567p}). Because our solar system likely formed in a clustered star-forming region \citep{adams2010birth,desch2024sun}, this similarity has been used to assert the claim that cometary water ice was directly inherited from the prestellar cloud or protostellar envelope with minimal chemical processing in the protoplanetary disk in which comets formed \citep{cleeves2014ancient}. In contrast, the gas HDO/H$_{2}$O ratios measured toward five distributed hot corinos found in isolated environments (i.e., they are not associated with star-forming cloud complexes) tend to be a factor of 2-10 higher \citep{jensen2019alma,jensen2021alma,andreu2023high,tobin2023deuterium}. Some of these studies suggested that such a difference in gas HDO/H$_{2}$O ratios could be due to the lower radiation experienced by more isolated star-forming regions, which implies colder and/or longer prestellar stages than those in more clustered star-forming regions. It must be emphasized, however, that this reported dependence of the protostellar gas HDO/H$_{2}$O on the local environment is based on a very limited sample size of sources, as discussed further in Section~\ref{subsec:hdo_h2o}.

Nevertheless, it is possible that the gas-phase water detected toward hot corinos has experienced some degree of chemical processing in the gas phase (e.g., isotopic exchange with H$_{2}$ gas), which would mean that the water accreted by bodies forming in protoplanetary disks could be substantially altered from its prestellar state. To ascertain that these gas HDO/H$_{2}$O ratios are in fact not significantly altered by gas-phase processing and therefore are representative of ice HDO/H$_{2}$O ratios, HDO ice must be quantified directly.

The advent of JWST, with its unprecedented sensitivity, offers an opportunity to revisit the search for HDO ice toward such star-forming regions. Recently, \cite{slavicinska2024jwst} demonstrated the capability of JWST to unambiguously detect HDO ice toward two massive protostars and subsequently calculated the ice HDO/H$_{2}$O ratios of their envelopes. Here, we focus on the source L1527 IRS (also IRAS 04368+2557, hereafter L1527), an isolated low-mass class 0/I protostar \citep{tobin2013modeling} found in the Taurus molecular cloud ($d$ = 140 pc) that is deeply embedded within a dense 0.9 M$_{\odot}$ envelope and has a 0.0075 M$_{\odot}$, 75-125 AU-sized disk observed in the edge-on configuration \citep{aso2017alma,kristensen2012water,tobin2013modeling}. The protostar currently has a mass around $\sim$0.3-0.5 M$_{\odot}$ \citep{aso2017alma,van2023early} but may eventually accrete enough material from its envelope to grow to a sun-like mass \citep{tobin20120}.

A low-resolution ($R$=$\lambda$/$\Delta\lambda$$\sim$100) IR spectrum of the ices toward L1527 from 2.5-5.0 $\mu$m was previously collected with the AKARI telescope. \cite{aikawa2012akari} analyzed the ice features in this spectrum and tentatively assigned a weak feature observed close to the noise level at $\sim$4.13 $\mu$m to HDO ice. While the 3 $\mu$m H$_{2}$O feature was saturated in this spectrum, an ice HDO/H$_{2}$O ratio was calculated by estimating the H$_{2}$O ice column density via fitting the wings of the 3 $\mu$m feature. The resulting ratio, 2.0$\times$10$^{-2}$, is notably higher than the subsequently derived hot inner gas HDO/H$_{2}$O ratios of other low-mass protostars \citep{persson2014deuterium,jensen2019alma,jensen2021alma}.

Here, we revisit the 4.1 $\mu$m spectral region of L1527 with the JWST Near-InfraRed Spectrograph (NIRSpec), whose sensitivity, spatial resolution, and spectral resolution far exceed those of the AKARI facility. Furthermore, we use the JWST Mid-InfraRed Instrument (MIRI) to expand the observed spectrum of L1527 to 27.9 $\mu$m, enabling the detection of the 12 $\mu$m H$_{2}$O ice libration feature. These new high-quality data allow us to determine a robust ice HDO/H$_{2}$O ratio within an ice envelope of a low-mass protostar and compare this value with gas HDO/H$_{2}$O ratios measured toward hot corinos. Such comparisons can serve to evaluate whether water in protostellar envelopes undergoes significant gas-phase reprocessing once it sublimates from ices, and whether such reprocessing affects the water that is eventually accreted by volatile-rich bodies formed in outer protoplanetary disks.

\section{Methods} \label{sec:methods}

\subsection{Observations} \label{subsec:observations}

\begin{figure}[h!]
\centering
\includegraphics[width=0.5\linewidth]{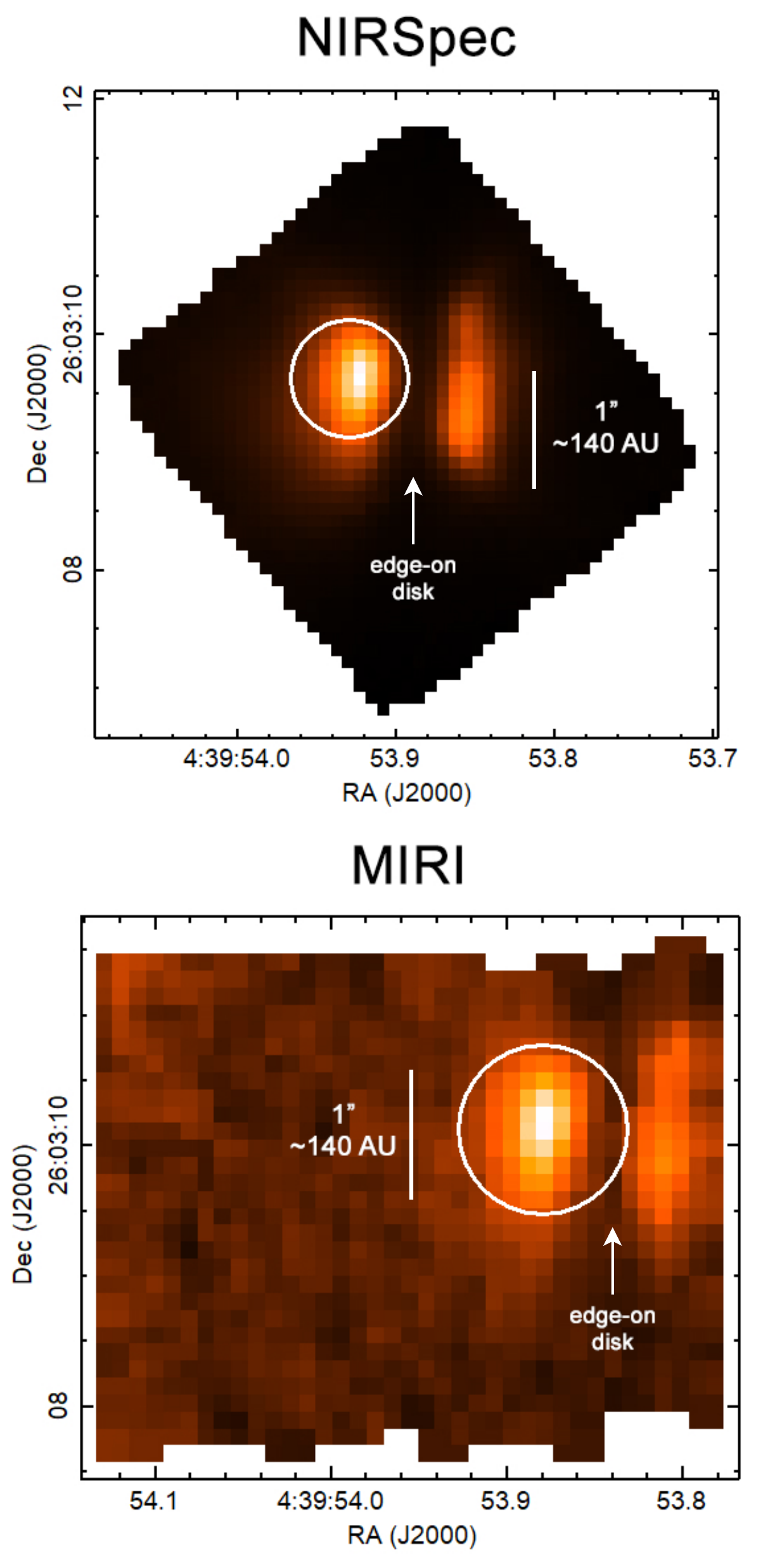}
\caption{5 $\mu$m continuum maps from the L1527 datacubes collected via NIRSpec IFU G395M (top) and MIRI-MRS (bottom) with the positions of the respective 1" and 1.30" diameter apertures used to extract the analyzed spectra shown in white.}
\label{fig:l1527_datacube}
\end{figure}

Medium-resolution ($R$=$\lambda$/$\Delta\lambda$$\sim$700-1300) IR spectra of L1527 covering the 4.1 $\mu$m HDO ice band were collected with the NIRSpec Integral Field Unit (IFU) using the G395M grating as part of the cycle 1 GO program \#1798 (PI John Tobin). NIRSpec IFU spectra collected with this grating provide continuous spectral coverage from 2.87-5.27 $\mu$m, unlike spectra collected with the G395H grating, which have a detector gap that overlaps with the 4.1 $\mu$m HDO ice feature. Additional low-resolution NIRSpec IFU PRISM ($R$=$\lambda$/$\Delta\lambda$$\sim$100) spectra from 0.6-5.3 $\mu$m collected as part of the cycle 1 GO program \#1960 (``JOYS+'', PI Ewine van Dishoeck), and MIRI Medium Resolution Spectrometer (MRS, $R$=$\lambda$/$\Delta\lambda$$\sim$1500-3500) spectra from 4.90-27.90 $\mu$m collected as part of the cycle 1 GTO program \#1290 (``JOYS'', PI Ewine van Dishoeck) gave continuous spectral coverage of L1527 from 0.6-27.9 $\mu$m. These JWST data can be accessed in the Mikulski Archive for Space Telescopes (MAST) via \dataset[doi: 10.17909/jhkb-w045]{https://doi.org/10.17909/jhkb-w045}.

The one-dimensional continuous 0.6-27.9 $\mu$m spectrum analyzed in this work was extracted using 1" and 1.3" diameter apertures for the NIRSpec and MIRI datacubes, respectively (Figure~\ref{fig:l1527_datacube}). The aperture sizes differ to account for the PSF difference between NIRSpec and MIRI at the wavelength where their spectral coverages overlap (MIRI PSF at 5 $\mu$m = $\sim$1.3$\times$ NIRSpec PSF at 5 $\mu$m, \citealt{law20233d,federman2024investigating}). The spectra did not require scaling for the NIRSpec and MIRI fluxes to align in the overlapping 4.90-5.27 $\mu$m wavelength range. The apertures were centered on positions that maximized the flux at 5 $\mu$m ($\alpha_{J2000}$=4:39:53.929, $\delta_{J2000}$=+26:03:09.617 in NIRSpec and $\alpha_{J2000}$=4:39:53.879, $\delta_{J2000}$=+26:03:10.119 in MIRI) to correct for small ($\sim$0.5") differences in coordinate calibration between NIRSpec and MIRI. Further details regarding the extraction position and data reduction can be found in Appendix~\ref{sec:app_obs}.

\subsection{Ice fitting} \label{subsec:analysis}
Although L1527 has a disk, its disk mass is $\gtrapprox$100 times less than the mass of its envelope -- i.e., the source is still relatively young and very deeply embedded. Therefore, unlike in class II sources where the ice envelopes have largely dissipated and most of any observed ice absorptions are expected to originate from disks, we expect ices in the envelope to dominate the observed ice absorption toward L1527, and that the ice absorptions observed at short (NIRSpec) and long (MIRI) wavelengths probe similar spatial regions of the protostellar system. To confirm that this assumption is reasonable, we used the diagnostic technique of comparing ice bands of the same species at different wavelengths suggested by \cite{pontoppidan2005ices}. Specifically, we compared the profiles and optical depths of the 4.39 $\mu$m $^{13}$CO$_{2}$ and 3.53 $\mu$m CH$_{3}$OH bands in the NIRSpec data with those of the 15.2 $\mu$m $^{12}$CO$_{2}$  and 9.7 $\mu$m CH$_{3}$OH bands in the MIRI data and found that the degrees of thermal processing and column densities indicated by these bands were consistent with each other (see Appendix~\ref{sec:nir_mir_comparison}). Thus, we opted to analyze the L1527 spectrum via classical techniques used to fit ice absorptions toward other class 0 and I sources (e.g., \citealt{gibb2004interstellar,boogert2008c2d}) rather than the complex radiative transfer modeling employed to analyze ice absorptions toward class II sources (e.g, \citealt{sturm2023jwst,bergner2024jwst}).

\subsubsection{HDO and CH$_{3}$OH ice fitting}

The 4 $\mu$m spectral region where the strongest HDO ice feature is found can also include contributions from CH$_{3}$OH and S-bearing ice species. This work used the procedures and laboratory data described in \cite{slavicinska2024jwst} to deconvolve these blended features and quantify the HDO and CH$_{3}$OH ice observed toward L1527 via the NIRSpec G395M data.

To briefly summarize this procedure, the CH$_{3}$OH ice column density was first obtained by fitting the 3.53 $\mu$m feature (isolated via a third-order polynomial local continuum subtraction, see Figure~\ref{fig:ch3oh_cont}) with CH$_{3}$OH ice laboratory spectra and a Gaussian representing NH$_{3}$ hydrates, although NH$_{4}$$^{+}$ salts can also have similar broad absorption feature around the same wavelength \citep{schutte2003origin}. More details on the CH$_{3}$OH ice fitting at 3.53 $\mu$m can be found in Appendix~\ref{subsec:ch3oh_ice}. The resulting CH$_{3}$OH ice column density, calculated using a band strength of 4.86$\times$10$^{-18}$ cm molec$^{-1}$ for the 3.53 $\mu$m band from \cite{luna2018densities}, was then used to constrain the expected contribution of the CH$_{3}$OH combination modes to the 3.7-4.2 $\mu$m spectral region (isolated via a fourth-order polynomial local continuum subtraction, see Figure~\ref{fig:hdo_cont}). The remaining spectral excess between 3.7-4.2 $\mu$m was subsequently fit with a combination of a Gaussian representing SH-bearing species and smoothed laboratory spectra of both amorphous and crystalline HDO ice. Band strengths of 6.9$\times$10$^{-17}$ and 6.8$\times$10$^{-17}$ cm molec$^{-1}$ from Table A2 in \cite{slavicinska2024jwst} corresponding to the temperatures of the best-fitting amorphous and crystalline HDO profiles, respectively, were used to calculate the total HDO ice column density. These corresponding temperatures should not be interpreted as directly representative of the ice temperatures, as this two-component fit is only a first-order approximation of the continuous thermal gradient that is expected along the line of sight. Details on the uncertainty calculations for the HDO and CH$_{3}$OH ice column density can be found in Appendix~\ref{sec:errors}.

\subsubsection{H$_{2}$O ice fitting} \label{subsubsec:h2o_analysis}

H$_{2}$O ice, the most abundant observable ice toward most protostars, has mid-IR absorption features at approximately 3, 4.5, 6, and 12 $\mu$m that vary significantly in strength \citep{mastrapa2009optical}. The 3 $\mu$m stretching feature is preferentially used to quantify H$_{2}$O ice whenever possible because out of the four main H$_{2}$O mid-IR features, it is the strongest and the least blended with other ice or dust bands. However, as a result of its strength and the relatively high abundance of water ice, it is frequently saturated toward the most embedded protostellar sight-lines, as is the case for L1527. Two alternative methods can be used to derive H$_{2}$O ice column densities when this is the case: 1) fitting one or both of the wings of the 3 $\mu$m feature (e.g., \citealt{aikawa2012akari}), or 2) fitting the 12 $\mu$m libration feature (e.g., \citealt{boogert2008c2d,chen2024joys+}). The first method is complicated by the fact that the wings of the 3 $\mu$m band shift with temperature, grain size, and chemical environment \citep{smith1989absorption,dartois2001search}, so H$_{2}$O ice column densities derived from this method are highly dependent on the physical and chemical parameters of the template H$_{2}$O ice spectra used in the fit, as discussed further in Appendix~\ref{sec:akari_comparison}. The second method requires full mid-IR spectral coverage due to the broad profile of the 12 $\mu$m feature as well as its blending with similarly broad neighboring silicate features at 10 and 18 $\mu$m.

In this work, we derived the H$_{2}$O ice column density by fitting the 12 $\mu$m band. We used the observed optical depths at 2.8, 4.5, and 6 $\mu$m as additional constraints (i.e., the blue wing of the 3 $\mu$m band and the 4.5 and 6 $\mu$m features should not be vastly overfit with the model). Details on the continuum fitting used to extract the spectrum into the optical depth scale and the silicate subtraction used to isolate the 12 $\mu$m band can be found in Appendix~\ref{sec:continua}. A set of H$_{2}$O ice spectral models with a range of temperatures and grain sizes were generated via the DHS model for irregular grains \citep{min2005modeling} with the \texttt{OPTOOL} code \citep{dominik2021optool}, using water ice optical constants from \citep{mastrapa2009optical}. The full continuum- and silicate-subtracted spectrum was then fit with various combinations of these models until a combination that met all of the fitting criteria (sufficiently fits the 12 $\mu$m band without vastly overfitting the blue wing of the 3 $\mu$m band and the 4.5 and 6 $\mu$m bands) was found. The final model used a power-law size distribution $N$(a) $\propto$ a$^{-3.5}$ with grain sizes ranging from 0.1-1.0 $\mu$m and included amorphous H$_{2}$O ice at 40 K, amorphous H$_{2}$O ice at 80 K, and crystalline H$_{2}$O ice at 80 K. Band strengths from \cite{mastrapa2009optical} corresponding to these respective ice temperatures and phases (2.3$\times$10$^{-17}$, 3.0$\times$10$^{-17}$, and 3.3$\times$10$^{-17}$ cm molec$^{-1}$) were then used to calculate the total H$_{2}$O ice column density by summing the contribution to the full 12 $\mu$m band from each of these fit components. As with the HDO, the temperatures used in this model should not be interpreted as extracted temperatures of the H$_{2}$O ice, as they are only a rough approximation of the expected continuous thermal gradient along the line of sight. Additionally, these fits use only pure H$_{2}$O ice laboratory data and thus do not account for possible changes in the peak profiles due to chemical effects (e.g., mixing with other ice species like CO$_{2}$, NH$_{3}$/NH$_{4}$$^{+}$, CH$_{3}$OH, etc.), which can be degenerate with temperature effects. Details on the uncertainty calculations for the H$_{2}$O ice column density can be found in Appendix~\ref{sec:errors}.

\section{Results and discussion} \label{sec:results}

\begin{figure}[t!]
\centering
\includegraphics[width=0.5\linewidth]{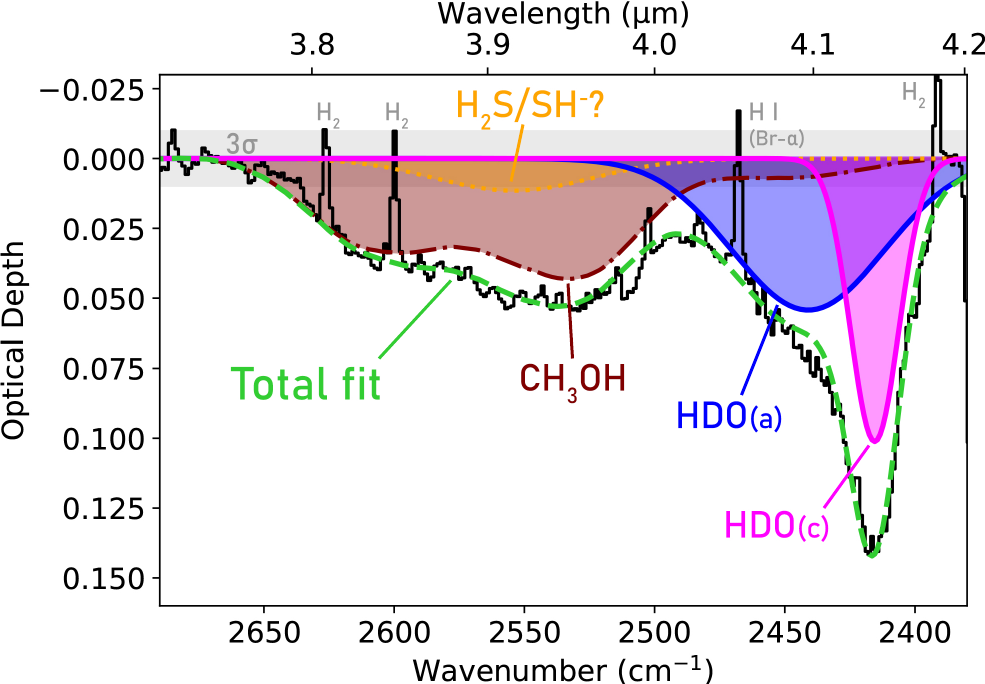}
\caption{NIRSpec spectrum of L1527 on the optical depth scale from 3.7-4.2 $\mu$m (solid black trace) along with the ice fit components CH$_{3}$OH (dash-dotted brown trace), a Gaussian representing SH-bearing ices (dotted yellow trace), amorphous HDO (solid blue trace), crystalline HDO (solid magenta trace), and the total combined ice fit (dashed green trace). The gray horizontal shaded region indicates the 3$\sigma$ noise level. See Figure~\ref{fig:hdo_cont} for the local continuum used to extract this spectrum.}
\label{fig:hdo_fit}
\end{figure}

\subsection{HDO ice} \label{subsec:hdo_ice}

Figure~\ref{fig:hdo_fit} presents the 3.7-4.2 $\mu$m region of the L1527 spectrum in optical depth units after the local continuum subtraction (see Appendix~\ref{sec:continua}). The CH$_{3}$OH column density from the 3.53 $\mu$m fit (see Appendix~\ref{subsec:ch3oh_ice}) applied to fit this spectral region leaves small ($\sim$0.01-0.02 $\tau$) and large ($\sim$0.14 $\tau$) excess absorptions between 3.9-4.0 and 4.0-4.2 $\mu$m, respectively. The large excess between 4.0-4.2 $\mu$m is observed throughout the entire datacube, including areas both with and without scattering wings around the 4.27 $\mu$m CO$_{2}$ band (see Figure~\ref{fig:hdo_cont}).

The absorption excess between 4.0-4.2 $\mu$m presents an asymmetric feature with a profile very similar to that observed by \cite{slavicinska2024jwst} toward higher mass sources, although with much higher optical depth and improved S/N. Given an RMS $\sigma$ value of $\sim$0.004 determined from the relatively featureless 3.71-3.77 $\mu$m range, the feature is detected at almost 40$\sigma$, far exceeding the S/N of the L1527 HDO ice detection reported by \cite{aikawa2012akari}. As with the massive protostellar spectra presented in \cite{slavicinska2024jwst}, the feature is fit well using a combination of amorphous and crystalline HDO ice profiles that correspond to temperatures of $\sim$100 and 70 K, respectively, in laboratory 0.4\% HDO:H$_{2}$O ice mixtures. Such a combination of amorphous and crystalline components approximates to a low degree a thermal gradient within the ice envelope, which is expected along the line of sight, and thus should not be interpreted as the extracted physical ice temperatures. The presence of both amorphous and crystalline HDO is consistent with additional spectral evidence for thermal processing toward this source, such as the double-peaked 4.39 $\mu$m $^{13}$CO$_{2}$ stretching and 15.2 $\mu$m $^{12}$CO$_{2}$ bending features \citep{pontoppidan2008c2d,brunken2024jwst}, and indicates that the most heated regions have experienced high enough temperatures to crystallize water ice. The HDO ice column density calculated from these fits, 9.2$^{+7.3}_{-1.7}$$\times$10$^{16}$ cm$^{-2}$ (see Table~\ref{tab:ice_columns}), is consistent with the L1527 HDO ice column density of 9.6$\times$10$^{16}$ cm$^{-2}$ reported from AKARI data by \cite{aikawa2012akari} despite the differences in the data quality and fitting methods (further discussion in Appendix~\ref{subsec:akari_hdo}).

The excess absorption between 3.9-4.0 $\mu$m is fit well with a Gaussian with a peak position of $\sim$3.9 $\mu$m and FWHM of $\sim$0.1 $\mu$m, which is similar to the profile of the S-H stretching mode of SH-bearing species like H$_{2}$S and NH$_{4}$SH in water-rich ices \citep{yarnall2022new,slavicinska2025ammonium}. The fit presented in Figure~\ref{fig:hdo_fit} results in column density upper limits of $\sim$0.3\% with respect to H$_{2}$O if assuming the only species contributing to the excess is H$_{2}$S and $\sim$13\% with respect to H$_{2}$O if assuming the only species contributing to the excess is NH$_{4}$SH, but these upper limits are extremely dependent on the continuum choice, resulting in estimated uncertainties on the order of $\sim$300\% and $\sim$70\%, respectively (see Appendix~\ref{sec:errors}).

\begin{figure}[h!]
\centering
\includegraphics[width=0.5\linewidth]{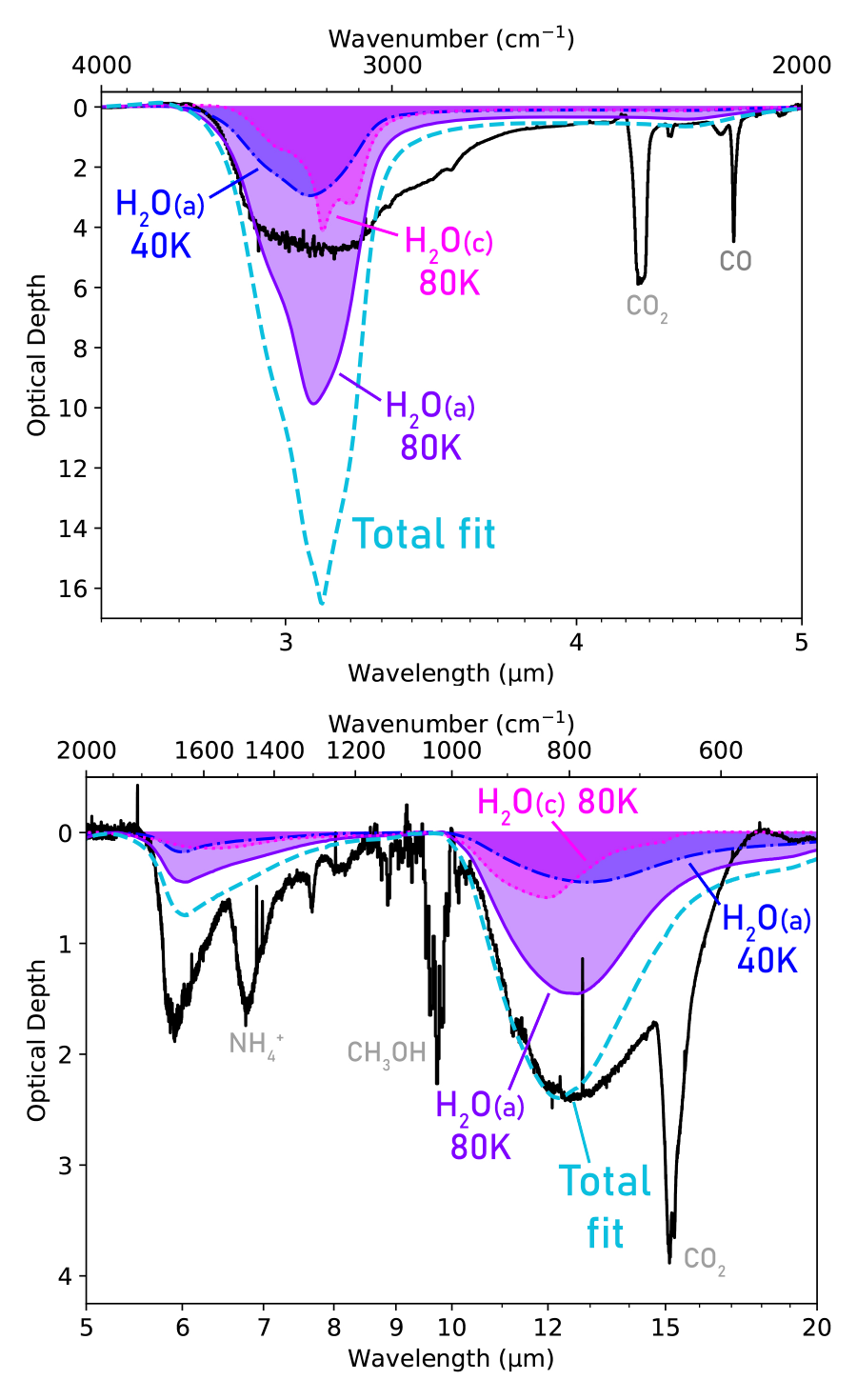}
\caption{Silicate-subtracted NIRSpec (top) and MIRI (bottom) spectra of L1527 on the optical depth scale (solid black trace) along with the fit components for H$_{2}$O ice: amorphous H$_{2}$O at 40 K (dash-dotted blue trace), amorphous H$_{2}$O at 80 K (solid purple trace), crystalline H$_{2}$O at 80 K (dotted magenta trace), and the total combined ice fit (dashed light blue trace). The observational data between 8.7-11.7 $\mu$m has been smoothed with a 10-point boxcar average to ameliorate the higher noise level due to the deep silicate feature at these wavelengths. Peaks of other major species (e.g., CO$_{2}$, CO) are labeled in gray but not fit. The temperatures indicated in this figure should not be interpreted as the extracted physical temperatures of the observed H$_{2}$O ice (see Sections~\ref{subsubsec:h2o_analysis} and~\ref{subsec:h2o_ice}). See Figure~\ref{fig:global_cont} for the global continuum fit and Figure~\ref{fig:silicate_sub} for the silicate subtraction.}
\label{fig:h2o_fit}
\end{figure}

\begin{table}[ht!]
\caption{Ice column densities and abundances toward L1527 and the optical depths of the features used to calculate them.}
\begin{center}
\begin{tabular}{c c c c c}
\hline
\hline
        Ice & Feature & Optical depth & Column density & Abundance\\
        & $\mu$m & $\tau$ & 10$^{17}$ cm$^{-2}$ & w.r.t. H$_{2}$O\\
        \hline
        H$_{2}$O & 12 & 2.0$^{+0.7}_{-0.4}$ & 209$^{+102}_{-51}$ & $\equiv$1 \\
        HDO & 4.1 & 0.14$^{+0.03}_{-0.01}$\tablenotemark{a} & 0.92$^{+0.73}_{-0.17}$ & 4.4$^{+3.7}_{-1.7}$$\times$10$^{-3}$ \\
        CH$_{3}$OH & 3.53 & 0.58$^{+0.05}_{-0.06}$\tablenotemark{b} & 22$^{+9}_{-11}$ & 0.11$^{+0.05}_{-0.06}$ \\
    \hline
\end{tabular}
\tablenotetext{a}{Optical depth of the entire amorphous and crystalline HDO absorption complex at 4.14 $\mu$m. The optical depths of the individual fitted amorphous and crystalline components are 0.05$^{+0.03}_{-0.01}$ and 0.101$^{+0.002}_{+0.005}$, respectively, with the asymmetric errors reflecting the differences in the optical depth of the fit components when using the ``upper and lower limit'' local continua shown in Appendix~\ref{sec:errors}.}
\tablenotetext{b}{Optical depth of the entire CH$_{3}$OH+NH$_{3}$ absorption complex at 3.53 $\mu$m.}

\label{tab:ice_columns}
\end{center}
\end{table}

\subsection{H$_{2}$O ice} \label{subsec:h2o_ice}

The profile of the isolated 12 $\mu$m H$_{2}$O libration band is fit with a combination of grain-shape corrected models of amorphous H$_{2}$O ice at 40 K, amorphous H$_{2}$O ice at 80 K, and crystalline H$_{2}$O ice at 80 K (Figure~\ref{fig:h2o_fit}). Such a combination of amorphous and crystalline components again approximates a thermal gradient within the ice envelope. This fit at 12 $\mu$m is consistent with the optical depth of the 4.5 $\mu$m combination mode and does not result in any overfitting of the 6 $\mu$m feature (which consists of the H$_{2}$O ice bending mode blended with other ice features, \citealt{boogert2008c2d}). However, the blue wing of the 3 $\mu$m band is slightly overfit, while the red wings of the 3 and 12 $\mu$m bands are underfit. It was not possible to obtain a satisfactory fit to these wings while still sufficiently fitting the 12 $\mu$m feature using only grain shape-corrected pure H$_{2}$O models. Including laboratory spectra of H$_{2}$O mixed with other major ice components (e.g., CO$_{2}$, NH$_{3}$, ammonium salts) may result in improvements to the fits in these wings, but we did not include such components in our fit for simplicity's sake, as deducing the chemical environment of H$_{2}$O ice is outside the scope of this work, and the inclusion of such components likely would not affect the final calculated H$_{2}$O ice column density outside of the reported uncertainties (see Appendix~\ref{sec:errors}) given that the band strength of the H$_{2}$O libration mode is not strongly affected by mixing with other common interstellar ice constituents \citep{kitta1983status}.

The H$_{2}$O ice column density obtained from the fit, 2.09$^{+1.02}_{-0.51}$$\times$10$^{19}$ cm$^{-2}$ (see Table~\ref{tab:ice_columns}), is a factor of 4 greater than the H$_{2}$O ice column density measured from the AKARI spectra \citep{aikawa2012akari}, which was likely underestimated due to the availability of only the wings of the 3 $\mu$m band for fitting and a single template H$_{2}$O ice spectrum at a very cold temperature (10 K) being used in the fit (see Appendix~\ref{subsec:akari_h2o}). A higher H$_{2}$O ice column density, on the order of 10$^{19}$ molec cm$^{-2}$, is expected toward this source given its massive envelope and very deep silicate feature (Figure~\ref{fig:jwst_vs_akari}). In fact, the high water column density and optical depth of the 9.7 $\mu$m silicate feature ($\tau$$\sim$4.2) of L1527 are typical of other deeply embedded low-mass Class 0 protostars now being probed by JWST (e.g., \citealt{chen2024joys+}).

While the 12 $\mu$m H$_{2}$O feature serves as a satisfactory means of measuring the total H$_{2}$O ice column density, it cannot reliably indicate ice physical properties in the way that the sharp, asymmetric profile of the 4.1 $\mu$m HDO band indicates crystallinity (Section~\ref{subsec:hdo_ice}) due to the degenerate effects of temperature, grain shape and size, chemical environment, and the choices in continuum and subtracted silicate template on its broad profile (e.g., \citealt{boogert2008c2d,mastrapa2009optical}). This means that the temperatures of the H$_{2}$O ice data used in the H$_{2}$O fit should not be interpreted as representative of the actual temperature of the observed H$_{2}$O ice, as possible effects of different chemical environments on the profile of the 12 $\mu$m band were not considered here (see Section~\ref{subsubsec:h2o_analysis}). The profile change of the 12 $\mu$m feature upon crystallization is also not as dramatic compared to the very evident sharpening of the 3 $\mu$m H$_{2}$O stretching mode, which is saturated toward this source. As mentioned in Section~\ref{subsubsec:h2o_analysis}, physicochemical effects on the wings of the 3 $\mu$m feature can be degenerate as well, so these wings are also unreliable tracers of physical properties when the 3 $\mu$m feature is saturated. For this reason, no secure conclusions can be drawn about the temperature or morphology of H$_{2}$O ice toward this source from this fit.

However, the total L1527 water ice abundance with respect to the hydrogen column density can provide other valuable information regarding the thermal processing history of the ice envelope. This value can be approximated using the relation between silicate optical depth and $N_{\rm{H}}$ in the dense cores of Lupus \citep{boogert2013infrared}. The resulting inferred H$_{2}$O ice abundance of L1527 is $\sim$4$\times$10$^{-5}$, on par with the higher end of H$_{2}$O ice abundance values measured toward other low-mass protostars (Figure~\ref{fig:jwst_vs_akari}). Such an abundance demonstrates that, although temperatures high enough to crystallize water ice must have been reached in at least some parts of the envelope of L1527 based on the 4.1 $\mu$m HDO ice band profile, a large portion of the envelope was likely not heated to high enough temperatures to desorb a significant portion of the H$_{2}$O ice ($>$100 K). The relatively blue peak position of the L1527 6.85 $\mu$m NH$_{4}$$^{+}$ feature, which is more redshifted toward sources that have lost most of their H$_{2}$O ice, is also consistent with this moderate processing scenario \citep{slavicinska2025ammonium}.

\subsection{HDO/H$_{2}$O ratio} \label{subsec:hdo_h2o}

\begin{figure}[h!]
\centering
\includegraphics[width=\linewidth]{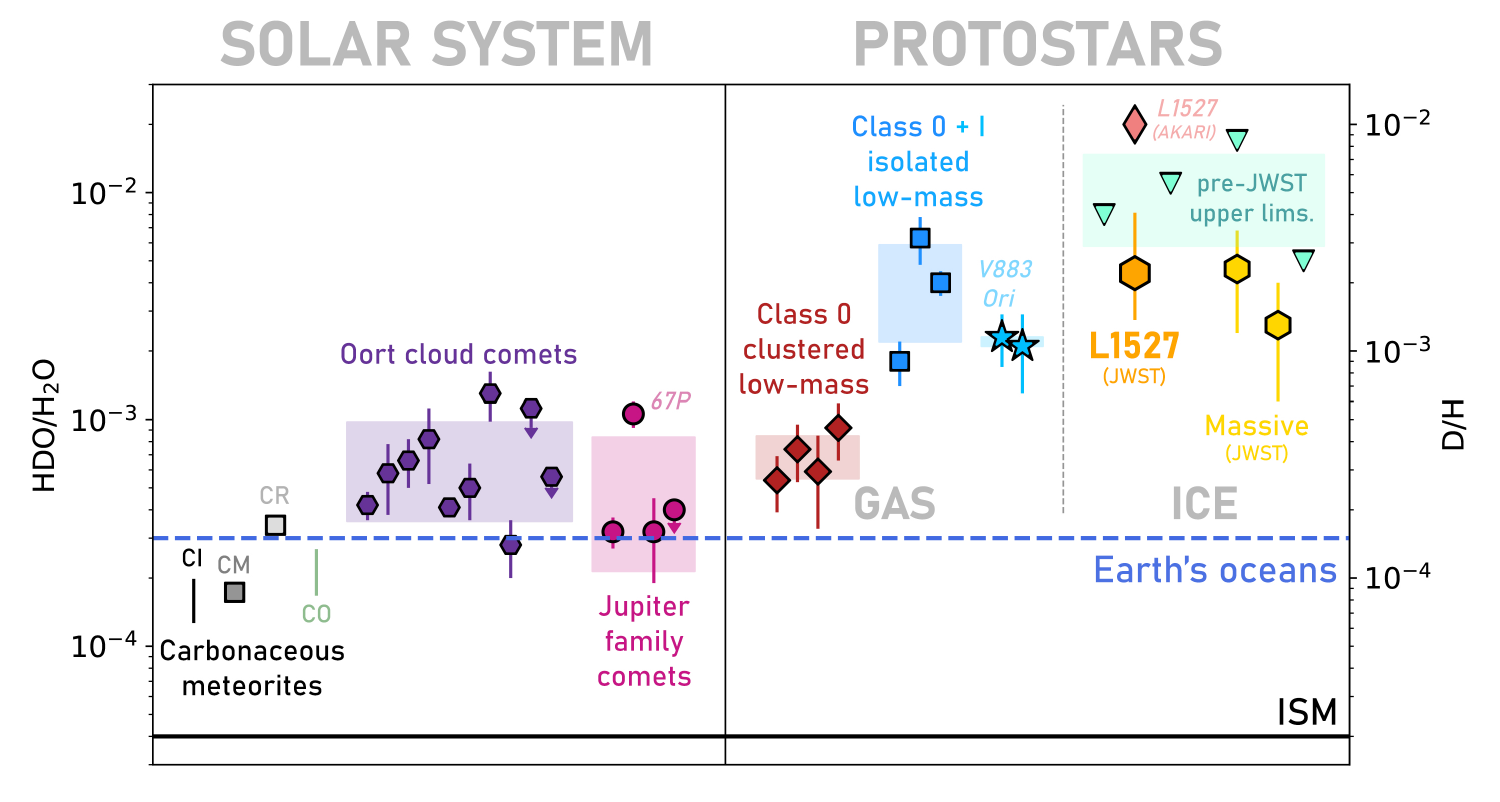}
\caption{Comparison of HDO/H$_{2}$O ratios measured in various primitive solar system objects and toward protostars in both ices and the gas phase. The ice HDO/H$_{2}$O ratio of L1527 measured in this work using JWST data is indicated with the large orange hexagon. The data and their sources can be found in Tables~\ref{tab:hdo_h2o_lit_ss} and~\ref{tab:hdo_h2o_lit_ps}.}
\label{fig:scatter_plot}
\end{figure}

The column densities of HDO and H$_{2}$O ice found here result in an ice HDO/H$_{2}$O ratio of 4.4$^{+3.7}_{-1.7}$$\times$10$^{-3}$. A comparison of this ratio with HDO/H$_{2}$O values and upper limits previously measured in protostellar envelopes as well as volatile-rich solar system objects is presented in Figure~\ref{fig:scatter_plot}.

Within the reported uncertainties, our value is consistent with the range of HDO/H$_{2}$O gas values that have been observed toward isolated low-mass protostars (1.8$\pm$0.4-6.3$\pm$1.5$\times$10$^{-3}$) in the hot inner regions of Class 0 and I protostellar envelopes \citep{jensen2019alma,jensen2021alma,andreu2023high}. This similarity supports the assumption that the gas HDO/H$_{2}$O ratios measured in the hot regions of protostars with sun-like masses are indeed representative of their ice HDO/H$_{2}$O ratios. Nevertheless, given the nearly half order of magnitude scatter observed in the gas HDO/H$_{2}$O ratios of isolated low-mass protostars, it is clear that parameters such as specific local conditions and evolutionary timescales may affect these ratios as well, so future work comparing gas and ice HDO/H$_{2}$O ratios toward the same protostar will be crucial for constraining the exact extent of gas-phase water reprocessing upon the sublimation of water ice in the envelopes of these objects.

\cite{tobin2023deuterium} recently showed that the gas HDO/H$_{2}$O ratio of 2.26$\pm$0.63$\times$10$^{-3}$ measured within the extended snowline of a disk of an isolated Class I protostar that recently experienced an accretion burst, V883 Ori, is generally consistent with the gas ratios measured in the hot inner regions of isolated Class 0 low-mass protostars. Such a consistency serves as evidence that the ices formed early in the star formation process, in prestellar clouds and protostellar envelopes, may be largely inherited by the outer regions of protoplanetary disks (and, by extension, the planetary bodies that form within them) of sun-like protostars without substantial chemical alteration via gas-phase reactions. Given that L1527 is an isolated low-mass protostar classified somewhere between the Class 0 and I stages \citep{tobin2013modeling}, the similarity of its ice HDO/H$_{2}$O ratio to gas HDO/H$_{2}$O ratios measured toward both Class 0 and Class I isolated sources, including V883 Ori, further corroborates this scenario.

The ice HDO/H$_{2}$O ratio of L1527 being consistent with the gas HDO/H$_{2}$O ratios of other isolated low-mass protostars means that it is also higher by a factor of $\sim$4-7 than the gas HDO/H$_{2}$O ratios measured in the hot inner regions of clustered Class 0 protostars and up to an order of magnitude higher than the HDO/H$_{2}$O ratios measured in comets (see Tables~\ref{tab:hdo_h2o_lit_ss} and~\ref{tab:hdo_h2o_lit_ps}). As mentioned previously, it has been suggested that water formed in isolated star-forming regions has higher D/H ratios than that formed in clustered regions because of either longer or colder prestellar stages, during which the majority of the ices in protostellar envelopes form \citep{jensen2019alma,jensen2021alma}. If this hypothesis is correct and local conditions greatly influence deuterium ratios, then our L1527 ice HDO/H$_{2}$O ratio is not fully comparable to cometary HDO/H$_{2}$O ratios because our Sun likely formed in a highly clustered and irradiated (and thus warmer) star-forming region \citep{adams2010birth,desch2024sun}. However, more samples are needed to ensure that this trend of higher HDO/H$_{2}$O ratios in more isolated protostars is statistically robust, as the current sample of protostellar gas HDO/H$_{2}$O values is still relatively small and consists of sources scattered throughout multiple molecular clouds in different galactic locations. The four currently published clustered protostellar gas HDO/H$_{2}$O values were measured toward the Ophiuchus and Perseus NGC 1333 molecular clouds, while the five isolated gas values were measured toward less populated regions in Orion and Taurus and three different isolated Bok globules. This means that any comparison of their values does not control for differences that could arise due to sampling from different star-forming regions. For this reason, larger sample sizes of both gas and ice HDO/H$_{2}$O measurements, particularly of protostars in the same molecular cloud but with different proximities to other stars, are crucial.

Furthermore, the ice HDO/H$_{2}$O ratio of L1527 is very similar to the ice HDO/H$_{2}$O ratios measured recently with NIRSpec IFU G395M toward more massive protostars (2.6$\pm$1.4 and 4.6$\pm$2.2 $\times$10$^{-3}$, \citealt{slavicinska2024jwst}). HOPS 370 in particular is located north of the extended Orion Nebula Cluster, a highly active star-forming region, so its similar ice HDO/H$_{2}$O ratio to L1527 is inconsistent with the above hypothesis. It is also inconsistent with the trend observed in the deuterium ratio of CH$_{3}$OH gas in protostars, which correlates negatively with stellar mass (i.e., more massive protostars tend to have lower CH$_{2}$DOH/CH$_{3}$OH ratios, \citealt{van2022methanol}). If isotopic ratios of species that formed primarily in the solid state do in fact reflect prestellar cloud conditions as has been previously suggested \citep{jensen2019alma,van2022methanol}, the similarity between HDO/H$_{2}$O ratios in low- and high-mass protostars, both in the ice as noted here and in some gas measurements as discussed previously in \cite{slavicinska2024jwst}, would imply similar prestellar conditions between isolated low-mass protostars and some massive protostars. However, the large scatter in the massive protostellar gas HDO/H$_{2}$O ratios and the small ice HDO/H$_{2}$O sample size in both isolated low-mass and massive protostars precludes drawing more general statements regarding the prestellar conditions of low-mass and massive protostars as a whole.

\section{Conclusions} \label{sec:concluzh}

We report the first JWST detection and quantification of HDO ice toward a low-mass protostar, L1527 IRS. The detection and quantification were performed using the 4.1 $\mu$m O-D stretching mode observed with the NIRSpec IFU instrument. Due to JWST's unprecedented sensitivity and spectral resolution, the detection exceeds the noise level by almost 40$\sigma$, and the feature is spectrally resolved enough to reveal an asymmetric profile characteristic of both amorphous and crystalline HDO ice in the envelope of L1527. We subsequently quantified the ice HDO/H$_{2}$O ratio toward this source using the 12 $\mu$m H$_{2}$O ice libration mode in the MIRI range. Our main conclusions can be summarized as follows:

\begin{enumerate}

\item We measured an ice HDO/H$_{2}$O ratio toward L1527 of 4.4$^{+3.7}_{-1.7}$$\times$10$^{-3}$, where the reported uncertainties are dominated by uncertainties in continuum definition and laboratory ice band strengths rather than observational uncertainties.

\item The L1527 ice HDO/H$_{2}$O ratio is similar to the gas HDO/H$_{2}$O ratios measured interferometrically toward the hot inner regions of other isolated low-mass protostars and within the snowline of the young protoplanetary disk V883 Ori. This similarity is in general agreement with the assumption that gas-phase molecular abundances in hot corinos and young disks are representative of those in protostellar ices, with minimal alteration caused by chemical processing by gas-phase reactions following ice sublimation.

\item The ice HDO/H$_{2}$O ratio of L1527 is a factor of 4-7 times greater than the gas HDO/H$_{2}$O ratios measured toward Class 0 clustered low-mass protostars and up to an order of magnitude greater than those measured in comets. This difference may be due to the water in these objects having undergone gas-phase reprocessing at some stage, or because of differences in the local star-forming environments being reflected in the HDO/H$_{2}$O ratio of ices formed during the prestellar stages. If the latter scenario is the case, then the ice HDO/H$_{2}$O ratio measured toward an isolated object like L1527 would be expected to exceed the ratios measured in comets in our solar system because our Sun likely formed in a more clustered environment.

\end{enumerate}

Moving forward, a comparison of the gas and ice HDO/H$_{2}$O ratios measured toward the same source will be essential to conclusively determine the extent of gas-phase chemical alteration of water in protostellar envelopes and protoplanetary disks given the large scatter in gas HDO/H$_{2}$O values measured in low-mass protostars. Additionally, measurements of HDO ice toward a larger sample of low-mass protostars in both clustered and isolated star-forming regions are needed to fully characterize the effects of local environmental conditions on ice HDO/H$_{2}$O ratios in protostellar envelopes and to conclusively evaluate the plausible link between protostellar and cometary water ice. The unprecedented sensitivity and spectral resolution of JWST opens the door for such measurements in the future.


\section{Acknowedgments}
K.S. thanks Brian Ferrari and Gary Huss for helpful scientific discussions. The authors thank an anonymous reviewer for their insightful comments that helped to improve this paper. This work is based on observations made with the NASA/ESA/CSA \textit{James Webb} Space Telescope. The data were obtained from the Mikulski Archive for Space Telescopes at the Space Telescope Science Institute, which is operated by the Association of Universities for Research in Astronomy, Inc., under NASA contract NAS 5-03127 for JWST. These observations are associated with programs \#1798, \#1290, and \#1960. All the data used in this paper can be found in MAST: \dataset[10.17909/jhkb-w045]{http://dx.doi.org/10.17909/jhkb-w045}. The following National and International Funding Agencies funded and supported the MIRI development: NASA; ESA; Belgian Science Policy Office (BELSPO); Centre Nationale d’Etudes Spatiales (CNES); Danish National Space Centre; Deutsches Zentrum fur Luft- und Raumfahrt (DLR); Enterprise Ireland; Ministerio De Economiá y Competividad; Netherlands Research School for Astronomy (NOVA); Netherlands Organisation for Scientific Research (NWO); Science and Technology Facilities Council; Swiss Space Office; Swedish National Space Agency; and UK Space Agency. Astrochemistry at Leiden is supported by funding from the European Research Council (ERC) under the European Union’s Horizon 2020 research and innovation programme (grant agreement No. 101019751 MOLDISK), the Netherlands Research School for Astronomy (NOVA), and the Danish National Research Foundation through the Center of Excellence ``InterCat'' (Grant agreement no.: DNRF150). A.C.G. acknowledges support from PRIN-MUR 2022 20228JPA3A ``The path to star and planet formation in the JWST era (PATH)'' funded by NextGeneration EU and by INAF-GoG 2022 ``NIR-dark Accretion Outbursts in Massive Young stellar objects (NAOMY)'' and Large Grant INAF 2022 ``YSOs Outflows, Disks and Accretion: towards a global framework for the evolution of planet forming systems (YODA)''.
 

%

\vspace{5mm}
\facilities{JWST (NIRSpec IFU and MIRI MRS)}


\software{\texttt{numpy} \citep{harris2020array}, \texttt{scipy} \citep{2020SciPy-NMeth}, \texttt{matplotlib} \citep{Hunter:2007}, \texttt{OPTOOL} \citep{dominik2021optool}, Adobe Photoshop}





\appendix

\counterwithin{figure}{section}
\counterwithin{table}{section}

\section{Observations} \label{sec:app_obs}

The extraction coordinates are centered on the upper layer of the L1527 disk seen in scattered light, slightly off the source (which lies in the dark area between the two scattered light lobes visible in Figure~\ref{fig:l1527_datacube}) to maximize S/N of the ice features. However, besides S/N differences, there is little variation in the profiles and relative optical depths of the ice features in the spectrum extracted from these coordinates and the spectrum extracted from the source center, and in fact, there are no noticeable spatial variations in the ice band profiles throughout most of the datacube.

The NIRSpec data were reduced via the JWST pipeline v1.16.0 using the Calibration Reference Data System
154 (CRDS) context file \texttt{jwst\_1298.pmap}. A wavelength offset of -0.004 $\mu$m was applied to the PRISM data to correct for what appears to be a discrepancy in wavelength calibration between PRISM and the G395M/H gratings observed in peak positions of both gas and ice lines. The MIRI data reduction is described in \cite{rangaswamy2025l1527}.

\section{Continuum, silicate, and CH$_{3}$OH ice fitting} \label{sec:continua}

\subsection{Local continuum at 3.7-4.2 $\mu$m}\label{subsec:hdo_cont}

\begin{figure}[h!]
\centering
\includegraphics[width=0.5\linewidth]{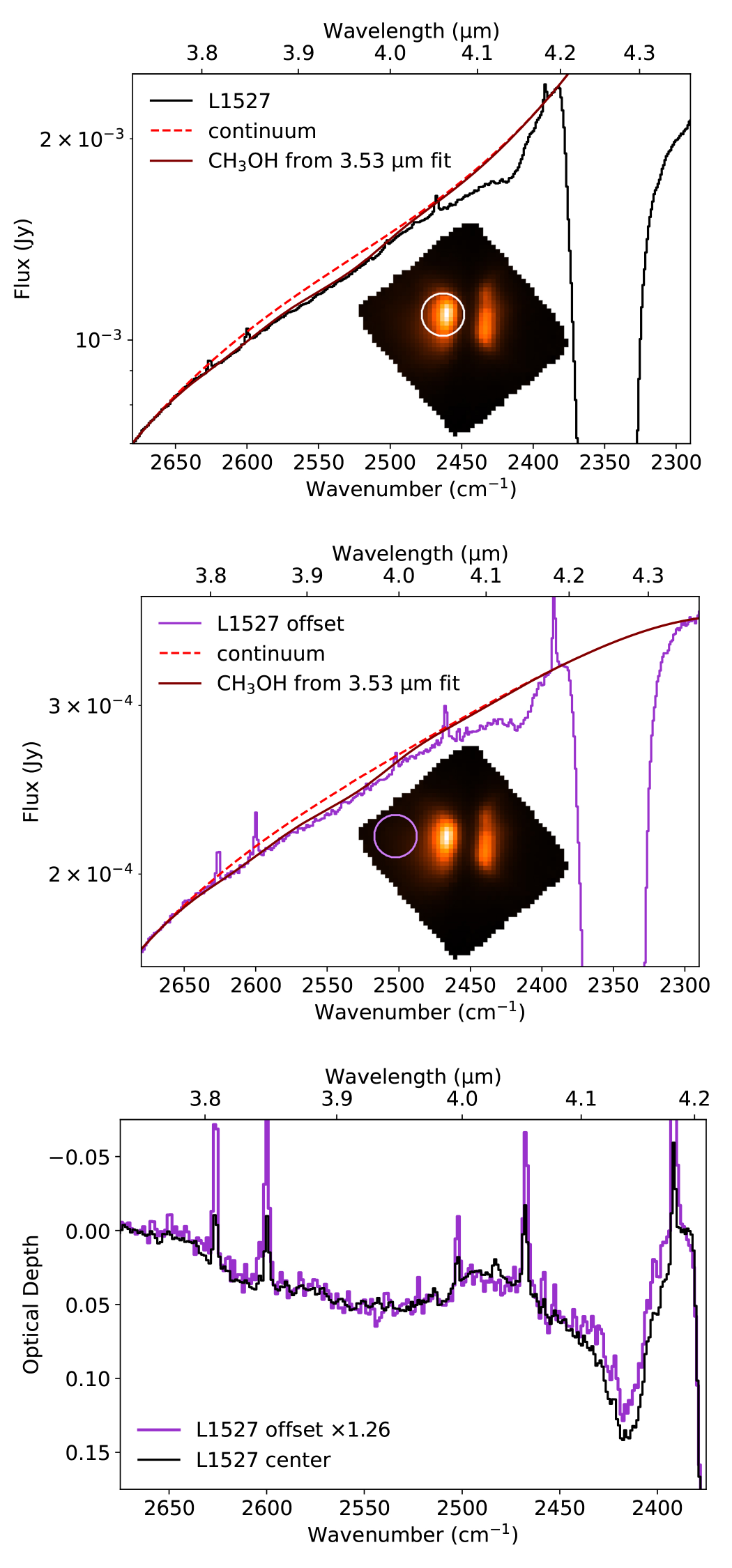}
\caption{NIRSpec spectra of L1527 from 3.7-4.2 $\mu$m extracted with 1" apertures from the region where the flux at 5 $\mu$m is maximized (top panel, black trace) and the region where ice absorption features are still present, but the CO$_{2}$ scattering wing is minimized (middle panel, purple trace). The dashed red traces indicate the local continua fit to these spectral regions, and the insets show the respective extracted regions in the NIRSpec datacube. The bottom panel compares the extracted optical depth spectra resulting from these local continuum fits to the two spectra. The local continuum in the top panel was used to calculate the optical depth spectrum presented in Figure~\ref{fig:hdo_fit}.}
\label{fig:hdo_cont}
\end{figure}

The top panel of Figure~\ref{fig:hdo_cont} shows the continuum applied to extract the optical depth spectrum of the 3.7-4.2 $\mu$m region that was used to calculate the HDO ice column density (Figure~\ref{fig:hdo_fit}). The local continuum fitting of this region is not trivial in this spectrum because the continuum is distorted by light scattering at the wings of the 4.27 $\mu$m CO$_{2}$ ice band, resulting in a break between the respective continua on the blue and red sides of the feature (e.g., \citealt{dartois2022influence}).

However, in other spatial regions within the datacube, data can be extracted where the CO$_{2}$ scattering wings are minimal, and such a break in the continuum around the feature does not occur. The middle panel of Figure~\ref{fig:hdo_cont} shows a spectrum extracted from such a region located east of the source. For this spectrum, the local continuum fitting at 3.7-4.2 $\mu$m is more straightforward as it can be defined so that it follows the flux on both sides of the 4.27 $\mu$m CO$_{2}$ band. This off-source spectrum can therefore be used to guide the local continuum fit to the higher S/N spectrum extracted from the region closer to the center of the source, where the continuum around the 4.27 $\mu$m band is disrupted by scattering.

The extracted spectra in optical depth resulting from these two local continuum fittings are shown in the bottom plot in Figure~\ref{fig:hdo_cont}. Although the local continua of the two spectra differ in shape, the profiles and relative optical depths of the extracted ice features are consistent with each other. The profiles and relative optical depths of ice features in spectral regions where local continuum fitting is more trivial because there are no spectrally adjacent strong features distorting the continuum (e.g., CH$_{3}$OH at $\sim$3.5 $\mu$m, $^{13}$CO$_{2}$ at $\sim$4.4 $\mu$m, and OCN$^{-}$ and CO at $\sim$4.6-4.7 $\mu$m) are also consistent, which shows that the ices observed along these two lines of sight have similar chemical compositions and have experienced similar degrees of thermal processing. However, the absolute optical depths of the ice features observed near the source center are on average a factor of $\sim$1.26 higher than those of the features in the offset position, which may be due to the ice envelope toward the offset position being thinner. For this reason, the offset spectrum in the bottom panel of Figure~\ref{fig:hdo_cont} is scaled by a factor of 1.26. The consistency between the extracted HDO absorption profiles provides confidence that the local continuum fit to the flux extracted from the central position sufficiently models the continuum distortion caused by the scattering wings of the 4.27 $\mu$m CO$_{2}$ ice band.

\subsection{Global continuum and silicate subtraction}

\begin{figure}[t!]
\centering
\includegraphics[width=0.5\linewidth]{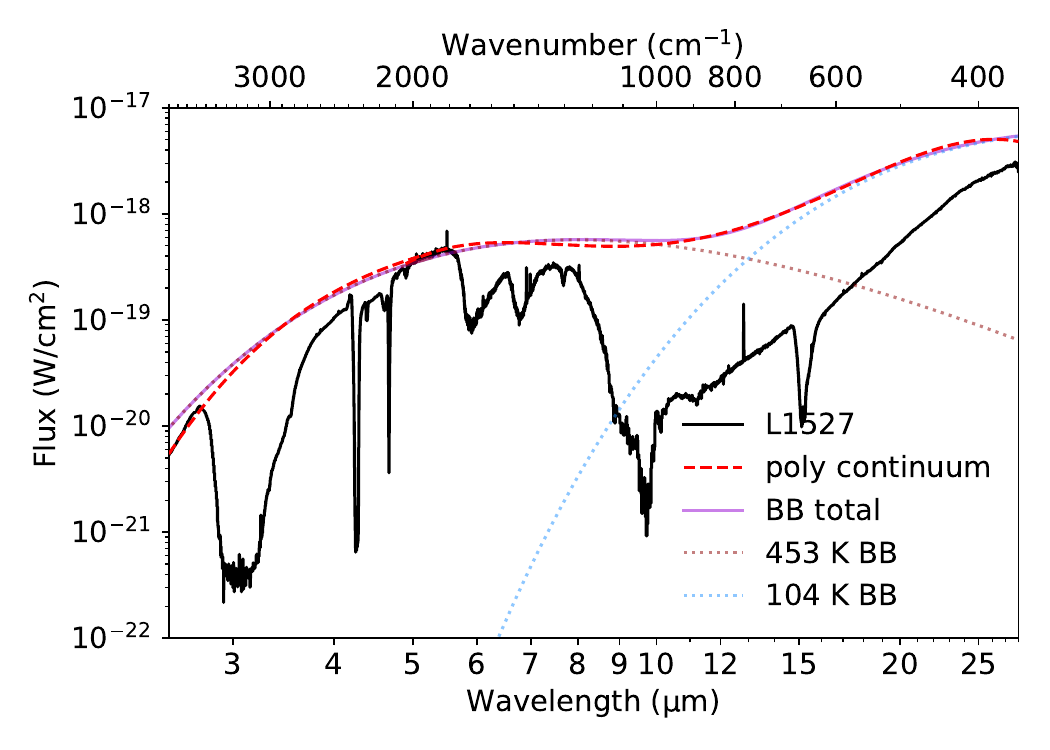}
\caption{Global continuum fit to the combined NIRSpec and MIRI spectrum of L1527 across the 2.5-27.9 $\mu$m spectral range. The observational data between 8.7-11.7 $\mu$m has been smoothed with a 10-point boxcar average to ameliorate the higher noise level due to the deep silicate feature at these wavelengths. The dashed red trace shows the polynomial continuum used to perform the global continuum subtraction, while the translucent purple trace shows how the profile of this polynomial can be reproduced with two blackbody emission profiles, one at 453 K (dotted red trace) and another at 104 K (dotted blue trace).}
\label{fig:global_cont}
\end{figure}

\begin{figure}[h!]
\centering
\includegraphics[width=0.5\linewidth]{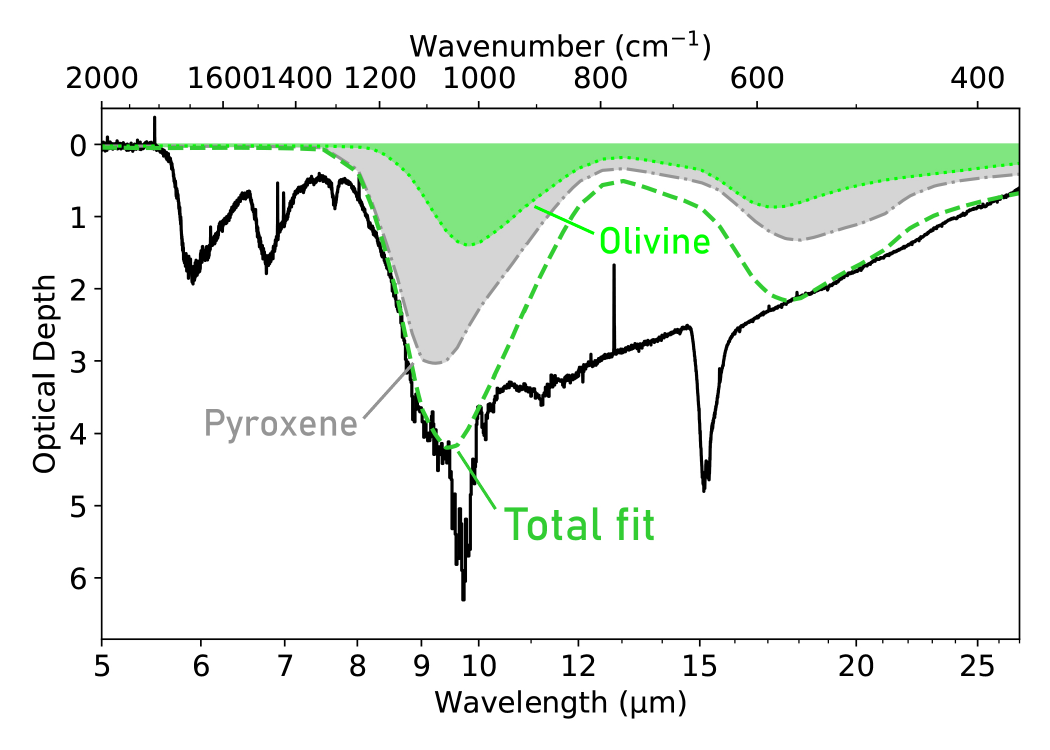}
\caption{MIRI spectrum of L1527 on the optical depth scale following global continuum subtraction (solid black trace) along with the silicate fit components pyroxene (dash-dotted gray trace) and olivine (dotted lime trace) and the total combined silicate fit (dashed green trace). This silicate fit was used to isolate the 12 $\mu$m H$_{2}$O ice libration feature shown in Figure~\ref{fig:h2o_fit}. The observational data between 8.7-11.7 $\mu$m has been smoothed with a 10-point boxcar average to ameliorate the higher noise level due to the deep silicate feature at these wavelengths.}
\label{fig:silicate_sub}
\end{figure}

To extract the full combined NIRSpec+MIRI spectrum into the optical depth scale, we defined two separate but connecting global continua in the near-IR (2.5-5 $\mu$m) and mid-IR (5-28 $\mu$m) ranges by fitting low-order polynomials to the data on a log($\lambda$F$_{\lambda}$) versus log($\lambda$) scale, similar to the procedure described in \cite{boogert2008c2d}. The resulting global continuum fit to the 2.5-27.9 $\mu$m spectral range is shown in Figure~\ref{fig:global_cont}. Similarly curved SEDs have been used to subtract the global continuum from other protostellar spectra (e.g., see the CrA IRAS 32, IRAS 03235+3004, and CRBR 2422.8-3423 global continua in Figure 1 of \citealt{boogert2008c2d}) and can be reproduced by radiative transfer models of protostellar systems \citep{crapsi2008characterizing}. In the case of our global continuum for L1527, the polynomial continuum fit can also be reproduced well with two blackbody curves, one at $\sim$450 K and another at $\sim$100 K, roughly simulating scattered light and thermal emission from different regions of the L1527 system.

We then isolated the 12 $\mu$m feature by subtracting a synthetic silicate spectrum generated for small spherical grains via the optical constants of amorphous olivine (MgFeSiO$_{4}$) and pyroxene (Mg$_{0.5}$Fe$_{0.5}$SiO$_{3}$) from \cite{dorschner1995steps}, as done previously for other dense and icy sightlines \citep{boogert2011ice}. The fit used for the silicate subtraction is shown in Figure~\ref{fig:silicate_sub}.

\subsection{Local continuum at 3.3-3.7 $\mu$m and CH$_{3}$OH ice fit} \label{subsec:ch3oh_ice}

\begin{figure}[h!]
\centering
\includegraphics[width=0.5\linewidth]{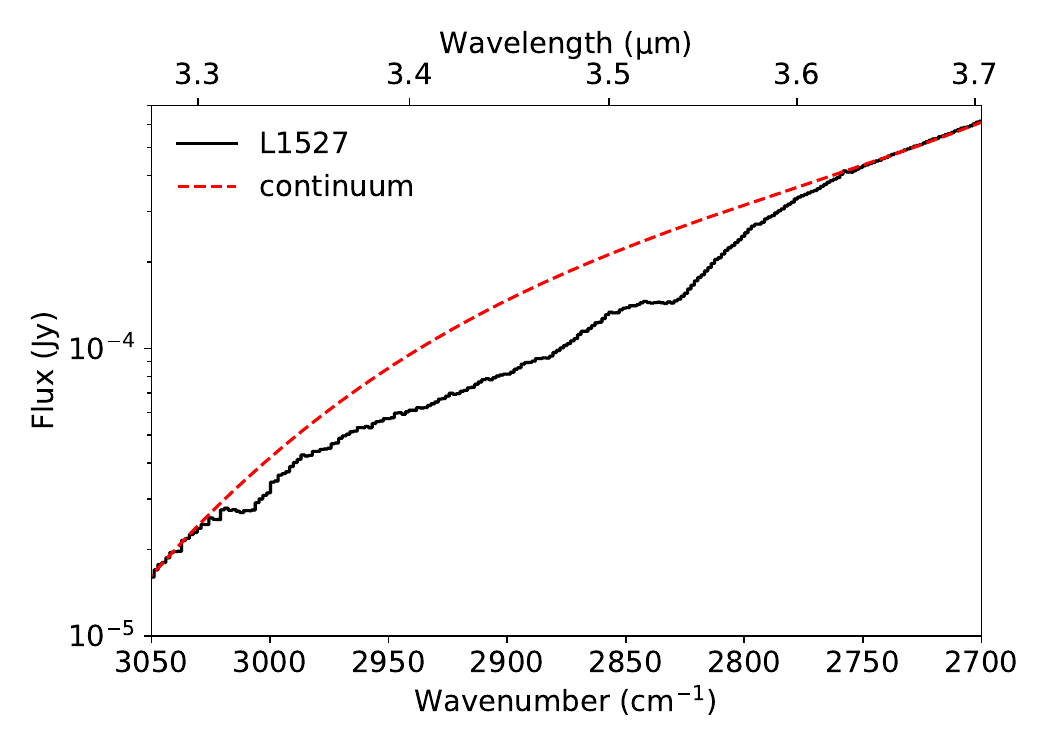}
\caption{Local continuum fit to the 3.3-3.7 $\mu$m spectral region used to extract the optical depth spectra in Figure~\ref{fig:ch3oh_fit}.}
\label{fig:ch3oh_cont}
\end{figure}

\begin{figure}[h!]
\centering
\includegraphics[width=0.5\linewidth]{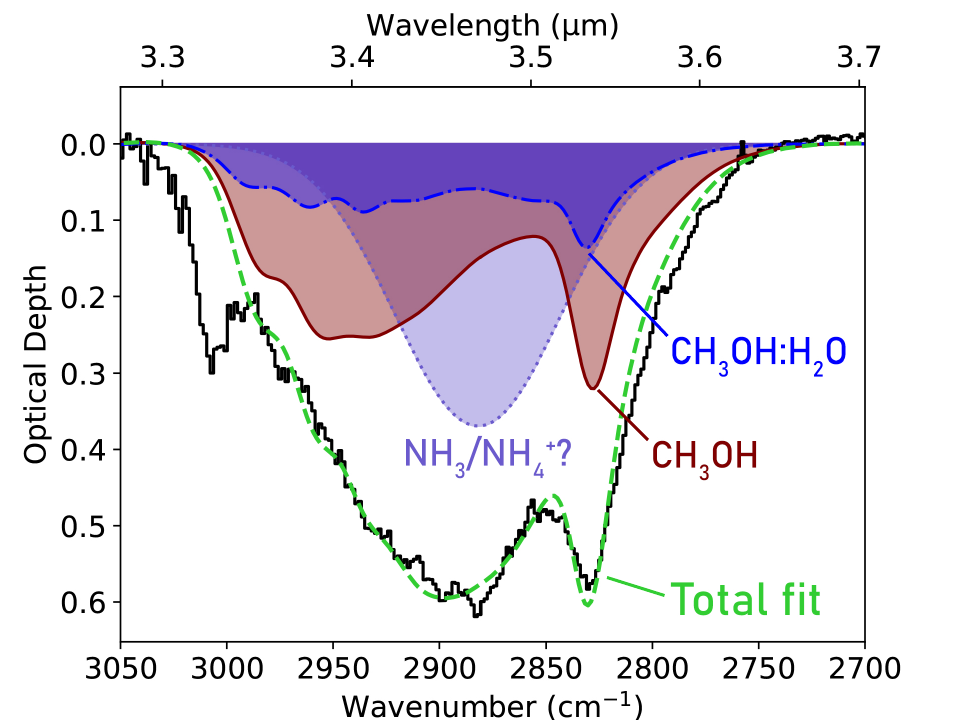}
\caption{NIRSpec spectrum of L1527 on the optical depth scale from 3.3-3.7 $\mu$m (solid black trace) along with the ice fit components pure CH$_{3}$OH (solid brown trace), a 1:5 CH$_{3}$OH:H$_{2}$O ice mixture (dash-dotted blue trace), a Gaussian representing NH$_{3}$ hydrates or NH$_{4}$$^{+}$ salts (dotted lavender trace), and the total combined ice fit (dashed green trace). This CH$_{3}$OH fit was used to constrain the contribution of CH$_{3}$OH to the 3.7-4.2 $\mu$m region shown in Figure~\ref{fig:hdo_fit}. See Figure~\ref{fig:ch3oh_cont} for the local continuum used to extract this spectrum.}
\label{fig:ch3oh_fit}
\end{figure}

The fit to the 3.3-3.7 $\mu$m region used to calculate the CH$_{3}$OH ice column density is shown in Figure~\ref{fig:ch3oh_fit}. Despite the obvious presence of thermally processed ices along the line of sight, the CH$_{3}$OH feature at 3.53 $\mu$m is best fit only with laboratory spectra of cold ($<$40 K) CH$_{3}$OH. Additionally, a combination of pure CH$_{3}$OH and CH$_{3}$OH in a water-dominated ice (in this case CH$_{3}$OH:H$_{2}$O 1:5) is needed to fit the feature. These fitting parameters are very similar to those used in \cite{slavicinska2024jwst} toward two massive protostars with thermally processed lines of sight. The resulting CH$_{3}$OH ice column density is 2.2$^{+0.8}_{-0.9}$$\times$10$^{18}$ cm$^{-2}$. The abundance of CH$_{3}$OH ice with respect to H$_{2}$O ice, $\sim$11\%, is close to the upper quartile value of CH$_{3}$OH ice abundances toward the sample of mostly Class I low-mass protostars observed before JWST \citep{boogert2015observations}.

\section{Comparing NIRSpec and MIRI bands of the same ice species} \label{sec:nir_mir_comparison}

\begin{figure}[h!]
\centering
\includegraphics[width=0.6\linewidth]{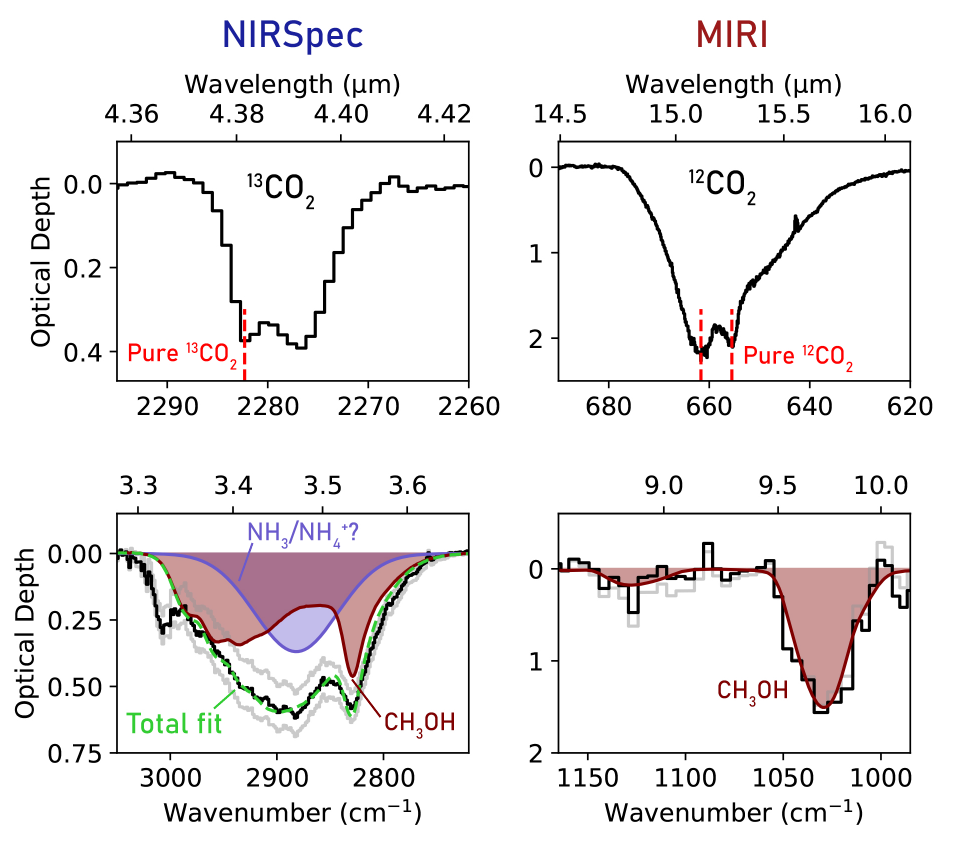}
\caption{Comparison of absorption bands from the same ice species at NIRSpec and MIRI wavelengths. Top left: the 4.39 $\mu$m $^{13}$CO$_{2}$ stretching mode (NIRSpec). Top right: the 15.2 $\mu$m $^{12}$CO$_{2}$ bending mode (MIRI). In both top plots, the red dashed lines mark peak positions indicating the presence of pure CO$_{2}$ ice that segregated from other ice species during thermal processing. Bottom left: fit to the 3.53 $\mu$m CH$_{3}$OH C-H stretching mode (NIRSpec). The gray traces indicate the optical depth of this spectral region extracted with the upper and lower limit continua shown in Figure~\ref{fig:ch3oh_cont_error}. Bottom right: fit to the 9.7 $\mu$m CH$_{3}$OH C-O stretching mode (MIRI) using the same combination of CH$_{3}$OH laboratory data used to fit the 3.5 $\mu$m complex in the bottom left panel. The data was binned using a binning factor of 30 to reduce the noise levels due to the CH$_{3}$OH band being located at the tip of the nearly saturated silicate feature. The black trace indicates the optical depth of this spectral region after the silicate subtraction shown in Figure~\ref{fig:silicate_sub}. The gray trace indicates the optical depth of this spectral region extracted using the local continuum fitting procedure described in Section 2.1 of \cite{bottinelli2010c2d}.}
\label{fig:nirspec_vs_miri_co2_ch3oh}
\end{figure}

Comparing the profiles and optical depths of bands found at different wavelengths but belonging to the same ice species can serve as a tool to evaluate whether disk radiative transfer effects need to be considered toward a source, as the ice column densities and chemical environments extracted from such bands of common species can differ in the same line of sight due to differences in photon paths through the disk at different wavelengths \citep{pontoppidan2005ices,sturm2023edge}. Such differences can indicate that the observed ice absorptions originate from ices located in different regions of the protostellar system (i.e., the short and long wavelengths do not probe the same spatial regions). Here we compare the $^{13}$CO$_{2}$ stretching mode at 4.39 $\mu$m and the $^{12}$CO$_{2}$ bending mode at 15.2 $\mu$m as well as the CH$_{3}$OH C-H stretching mode at 3.53 $\mu$m and the CH$_{3}$OH C-O stretching mode at 9.7 $\mu$m to investigate if the ice column densities and physicochemical environments extracted from these profiles at different wavelengths are consistent with each other, as would be expected for ice absorptions that originate from an isotropic envelope made up of small ($\lessapprox$1 $\mu$m) grains, where the same complex radiative transfer effects are not at play.

Figure~\ref{fig:nirspec_vs_miri_co2_ch3oh} shows the comparison of these bands. First, both the 4.39 $\mu$m $^{13}$CO$_{2}$ feature and the 15.2 $\mu$m $^{12}$CO$_{2}$ feature have double-peaked profiles indicative of ice thermal processing (see \citealt{boogert2000iso,pontoppidan2008c2d,brunken2024jwst}). This consistency demonstrates that the degrees of thermal processing experienced by the ices probed at short and long wavelengths are similar. Furthermore, the $^{12}$C/$^{13}$C ratio derived from these two bands by \cite{brunken2024joys+} for L1527, 105, is consistent with the range of $^{12}$C/$^{13}$C ratios derived for less evolved class 0 sources with split CO$_{2}$ features ($\sim$100-120). Although the 4.27 $\mu$m $^{12}$CO$_{2}$ stretching feature is saturated toward L1527 and the narrow 2.70 $\mu$m $^{12}$CO$_{2}$ combination mode is not resolved well enough in the low-resolution PRISM spectrum to derive a $^{12}$CO$_{2}$ column density from the shorter wavelengths, \cite{brunken2024joys+} found that for similar sources with similar $^{12}$C/$^{13}$C ratios and coherent 2.70, 4.27, and 15.2 $\mu$m bands, the $^{12}$CO$_{2}$ column densities derived from NIRSpec and MIRI data are remarkably consistent (typically within $\sim$5-20\% error).

The CH$_{3}$OH ice column density derived from fitting the 9.7 $\mu$m band using the same combination of pure CH$_{3}$OH and CH$_{3}$OH:H$_{2}$O spectra used to fit the 3.5 $\mu$m absorption complex results in a CH$_{3}$OH column density of 3.1$\pm$0.6$\times$10$^{18}$ molec cm$^{-2}$ using a band strength of 1.56$\times$10$^{-17}$ cm molec$^{-1}$ from \cite{luna2018densities}. This CH$_{3}$OH column density is consistent with the CH$_{3}$OH column density derived from the 3.53 $\mu$m band within the reported uncertainties that take into account both uncertainties in the continuum subtraction and band strengths. The difference between the two values, $\sim$33\%, is similar to the difference between CH$_{3}$OH column densities from the 3.53 and 9.7 $\mu$m bands reported for observations of dense clouds \citep{mcclure2023ice}.

These consistencies in the ice morphologies and column densities between ice bands of the same species probed at different wavelengths support the notion that the short and long wavelengths probe similar spatial regions in the L1527 protostellar system and provide confidence that the ice absorptions observed toward L1527 can be analyzed using classical ice fitting techniques.

\section{Ice fitting uncertainties} \label{sec:errors}

\begin{figure}[t!]
\centering
\includegraphics[width=0.7\linewidth]{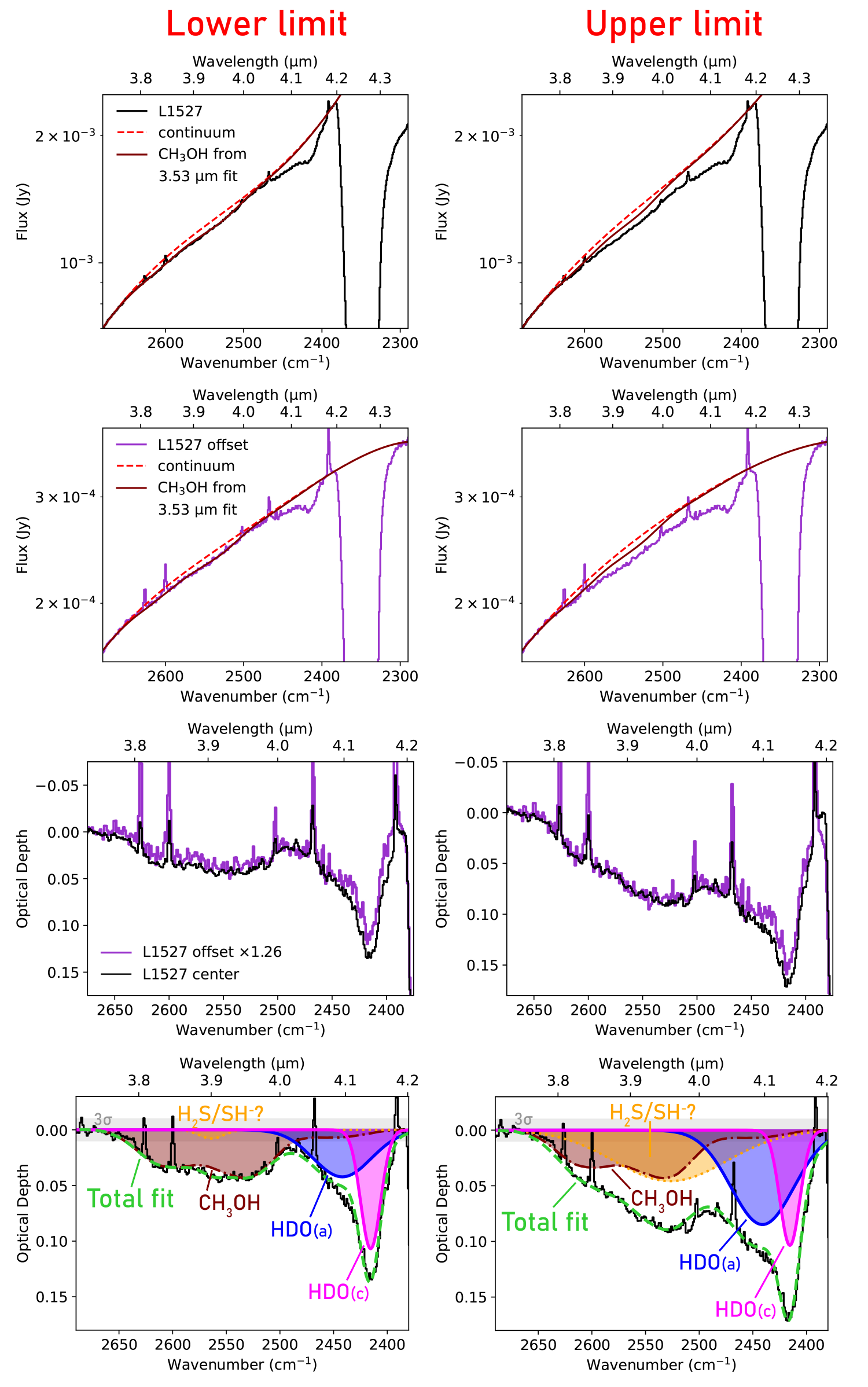}
\caption{The local continua and subsequent fits to the 3.7-4.2 $\mu$m region used to define continuum uncertainties on the HDO column densities. First row: the lower (left) and upper (right) limits on the local continua in the central spectrum. Second row: the corresponding lower (left) and upper (right) limits on the local continua in the off-source spectrum. Third row: comparisons of the extracted spectra resulting from the lower (left) and upper (right) limit continua between the two positions. Fourth row: ice fits to the spectra extracted using the continua in the top row. The resulting HDO column densities were used to define the HDO column density uncertainty from the continuum choice.}
\label{fig:hdo_cont_error}
\end{figure}

\begin{figure}[t!]
\centering
\includegraphics[width=0.7\linewidth]{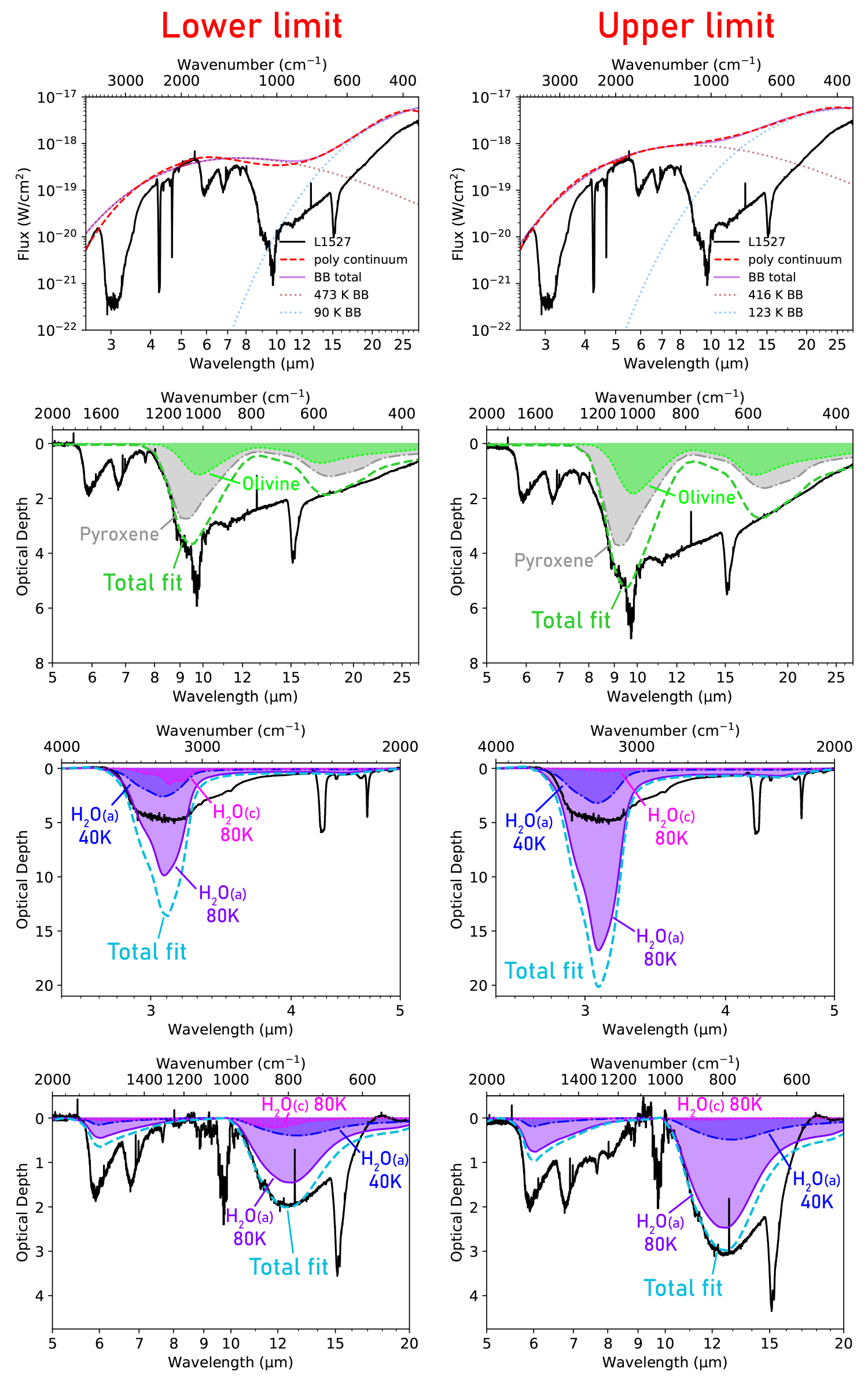}
\caption{The global continua and subsequent silicate subtractions and H$_{2}$O ice fits used to define continuum uncertainties on the H$_{2}$O column densities. First row: the lower (left) and upper (right) limits on the global continua in the central spectrum. Second row: the silicate subtraction performed on the optical depth-scale spectra extracted using the continua in the top row. Third and fourth row: H$_{2}$O ice fits to the NIRSpec and silicate-subtracted MIRI spectra. The resulting H$_{2}$O column densities were used to define the H$_{2}$O column density uncertainty from the continuum choice.}
\label{fig:h2o_cont_error}
\end{figure}

\begin{figure}[h]
\centering
\includegraphics[width=0.7\linewidth]{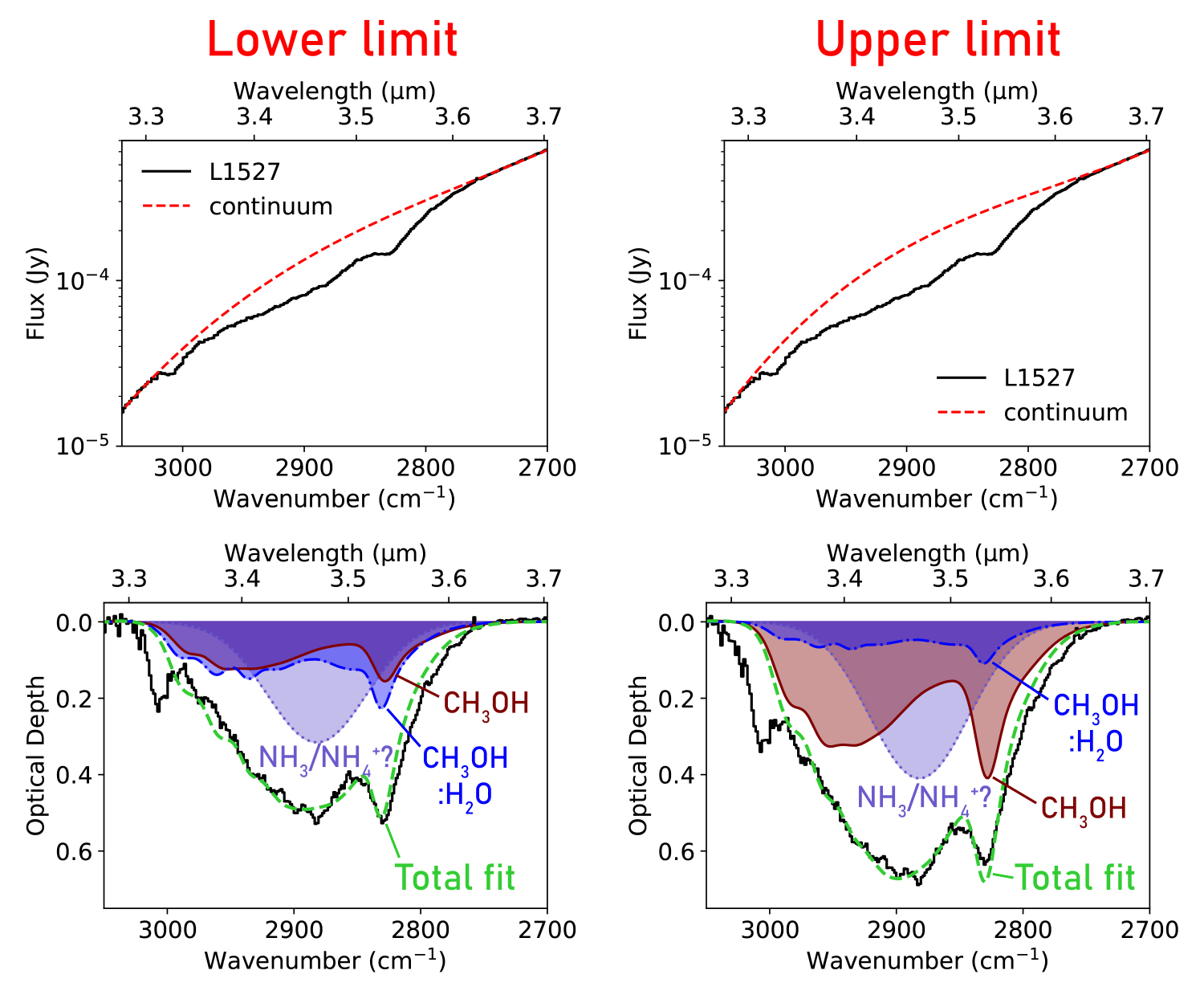}
\caption{The local continua and subsequent fits to the 3.3-3.7 $\mu$m region used to define continuum uncertainties on the CH$_{3}$OH column densities. Top row: the lower (left) and upper (right) limits on the local continua in the central spectrum. Bottom row: ice fits to the spectra extracted using the continua in the top row. The resulting CH$_{3}$OH column densities were used to define the CH$_{3}$OH column density uncertainty from the continuum choice.}
\label{fig:ch3oh_cont_error}
\end{figure}

The reported uncertainties on the HDO, H$_{2}$O, and CH$_{3}$OH ice column densities were determined here using methods similar to those described in \cite{slavicinska2024jwst}. Two main sources of uncertainty were considered: uncertainty of the local/global continuum choice, and uncertainty of the laboratory band strengths used to calculate the ice column densities. Compared to these two sources of uncertainty, other sources of uncertainty like observational noise and fitting covariances are small and are therefore treated here as negligible.

We quantified the uncertainties on the column densities from continuum uncertainty by defining the lower and upper limits of what we consider to be a ``reasonable'' continuum (top panels of Figures~\ref{fig:hdo_cont_error}, ~\ref{fig:h2o_cont_error}, and ~\ref{fig:ch3oh_cont_error}) and then using the same methods described in Section~\ref{sec:methods} to derive lower and upper limits of HDO, H$_{2}$O, and CH$_{3}$OH column densities by fitting the optical depth spectra extracted via these continua (bottom panels of Figures~\ref{fig:hdo_cont_error}, ~\ref{fig:h2o_cont_error}, and ~\ref{fig:ch3oh_cont_error}). In the case of HDO, we also ensured that a similar extracted spectral profile could be achieved with a ``reasonable'' local continuum fit to the spectrum extracted from the off-center position discussed in Section~\ref{subsec:hdo_cont} (second and third rows in Figure~\ref{fig:hdo_cont_error}). These column density lower and upper limits serve as estimates of the column density uncertainties from uncertainties in continuum fitting. They are asymmetric and, to an extent, subjective depending on what one considers to be a ``reasonable'' continuum.

The reported laboratory band strengths of the HDO and H$_{2}$O bands used here to calculate the ice column densities can vary significantly with temperature, particularly for the amorphous ices. Because we do not attempt to model the temperature gradient with our fits, we treat this variation of band strength with temperature as an additional source of uncertainty in the ice column densities. We again calculate lower and upper limits of HDO and H$_{2}$O ice column densities, this time by using the respective highest and lowest band strength measured for a given phase of the ice. The resulting uncertainties are again asymmetric. For CH$_{3}$OH, temperature and chemical environment changes can change the band strength of the 3.53 $\mu$m feature by a maximum of $\sim$20\% \citep{kerkhof1999infrared,luna2018densities}, and \cite{luna2018densities} report experimental uncertainties of 20\%. For all three ice species, the band strength and continuum uncertainties are both considered systematic and are thus added linearly to provide the total ice column density uncertainties.

Although SH-bearing ice species are not the focus of this work, we report in Section~\ref{subsec:hdo_ice} H$_{2}$S and NH$_{4}$SH upper limits ($\sim$0.3 and 13\% w.r.t. H$_{2}$O, respectively) calculated from the Gaussian fit to the excess at 3.9 $\mu$m, in the case that these upper limits may be useful for other studies of sulfur in star-forming regions or interstellar ices. However, it is evident from Figure~\ref{fig:hdo_cont_error} that the optical depth of this Gaussian can vary dramatically with the continuum choice, resulting in high uncertainties in these reported upper limits. Depending on the continuum, the H$_{2}$S upper limits can range from 0.07-2\%. For NH$_{4}$SH, the upper limit is $\sim$4\% in the case of the lower limit continuum choice shown in the bottom left panel of Figure~\ref{fig:hdo_cont_error}. In the case of the upper limit continuum choice in the bottom right panel, the 3.9 $\mu$m Gaussian profile is too broad to be attributed exclusively with NH$_{4}$SH. Furthermore, the L1527 NH$_{4}$$^{+}$ column density derived from the 6.85 $\mu$m feature \citep{slavicinska2025ammonium} provides an additional constraint that the NH$_{4}$SH upper limit cannot exceed that derived from the fit presented in Figure~\ref{fig:hdo_fit} ($\sim$13\%).

\section{Comparison with AKARI results} \label{sec:akari_comparison}

\subsection{HDO ice} \label{subsec:akari_hdo}

While \cite{aikawa2012akari} cautioned that their HDO column densities obtained from AKARI data toward L1527 ``should be interpreted with caution'' given the low signal-to-noise (``small bumps and hollows that deviate from the HDO feature''), the HDO ice column density measured here is similar to their value. Although the AKARI data were not spectrally resolved enough to reveal the asymmetric profile of the 4.1 $\mu$m band and were thus only fit with a single Gaussian, \cite{aikawa2012akari} noted that the profile of their best-fitting Gaussian was narrower and peaked at longer wavelengths than expected for cold and amorphous HDO ice and thus stated that the feature resembles HDO ice that has been annealed. While this is in contradiction with their fitting of 10 K amorphous H$_{2}$O ice to the 3 $\mu$m band wings, it is in agreement with our analysis that includes a prominent crystalline component of the HDO band.

\subsection{H$_{2}$O ice} \label{subsec:akari_h2o}

\begin{figure}[h!]
\centering
\includegraphics[width=0.5\linewidth]{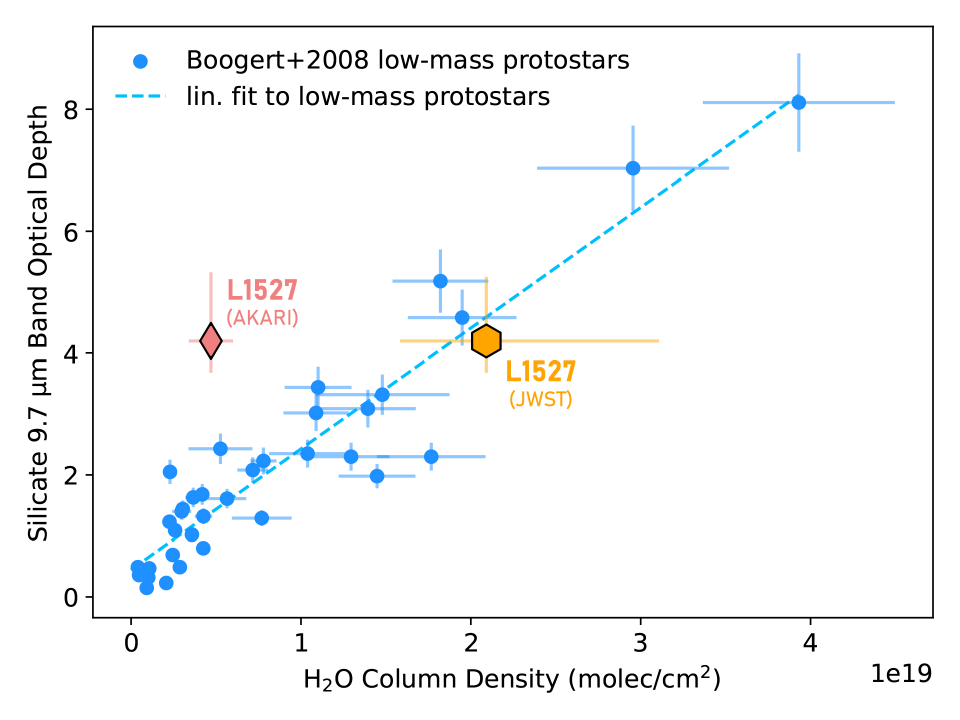}
\caption{Optical depth of the 9.7 $\mu$m silicate feature plotted as a function of H$_{2}$O ice column density toward low-mass protostars. The blue data points were obtained from Spitzer data \citep{boogert2008c2d}, and the blue dashed line shows the best linear fit to the data. The red diamond represents L1527 using the H$_{2}$O ice column density from \cite{aikawa2012akari}, while the golden hexagon represents L1527 using the H$_{2}$O ice column density from this work. Both L1527 data points use the silicate optical depth obtained from the MIRI spectrum in this paper.}
\label{fig:jwst_vs_akari}
\end{figure}

Our H$_{2}$O ice column density, 2.09$^{+1.02}_{-0.51}$$\times$10$^{19}$ cm$^{-2}$, is over a factor of 4 greater than the H$_{2}$O ice column density measured toward L1527 using AKARI spectra by \cite{aikawa2012akari}, 4.7$\pm$1.3$\times$10$^{18}$ cm$^{-2}$. This is because \cite{aikawa2012akari} derived their H$_{2}$O ice column densities by fitting the wings of the 3 $\mu$m feature with pure, 10 K H$_{2}$O ice spectra. Because the wings of this feature can redshift significantly with grain effects and higher temperatures, it is easy to over- or underfit this feature when fitting only its wings. \cite{aikawa2012akari} state that ``the H$_{2}$O ice column density cannot be significantly larger than we estimated'' because their fit to the 3 $\mu$m band wings is also consistent with their extracted optical depth of the 4.5 $\mu$m combination mode, but the optical depth of this extremely broad feature can change significantly with even small changes in the fitted global continuum. In past literature, the continuum from 2-5 $\mu$m has frequently been defined using datapoints around 4 $\mu$m (e.g., \citealt{gibb2004interstellar}) despite the fact that numerous weak ice features, including the broad blue wing of the H$_{2}$O combination mode, have absorptions in this spectral region. Such fitting can result in an underestimation of the H$_{2}$O ice combination mode optical depth.

Further supporting our significantly larger L1527 H$_{2}$O ice column density is a comparison of the 9.7 $\mu$m silicate feature optical depth and the H$_{2}$O ice column density observed toward L1527 versus those of other low-mass protostars (Figure~\ref{fig:jwst_vs_akari}). In general, the optical depth of the 9.7 $\mu$m feature strongly correlates with the observed H$_{2}$O column density toward these objects, and based on the high 9.7 $\mu$m feature optical depth observed toward L1527 ($\tau$ $\sim$4.2), its H$_{2}$O ice column density is indeed expected to be $\gtrsim$2$\times$10$^{19}$ cm$^{-2}$.

\section{Literature HDO/H$_{2}$O values}

\begin{table}[ht!]
\caption{Overview of HDO/H$_{2}$O values of various solar system objects.}
\begin{center}
\begin{tabular}{c c c c}
\hline
\hline
        Object & Phase & HDO/H$_{2}$O & Ref.\\
        & & 10$^{-3}$ & \\
        \hline
        \multicolumn{4}{c}{Carbonaceous Chondrites} \\
        \hline
        CI chondrites & - & 0.129-0.195 & 1 \\
        CM chondrites & - & 0.173$\pm$0.0072 & 1 \\
        CR chondrites & - & 0.342$^{+0.034}_{-0.02}$ & 1 \\
        CO chondrites & - & 0.170-0.264 & 1 \\
        \hline
        \multicolumn{4}{c}{Oort Cloud Comets} \\
        \hline
        1P/Halley & gas & 0.42$\pm$0.06 & 2 \\
        C/1996 B2 Hyatuake & gas & 0.58$\pm$0.20 & 3 \\
        C/1995 O1 Hale-Bopp & gas & 0.66$\pm$0.16 & 4 \\
        8P/Tuttle & gas & 0.82$\pm$0.29 & 5 \\
        C/2009 P1 Garradd & gas & 0.41$\pm$0.04 & 6 \\
        C/2002 T7 LINEAR & gas & 0.50$\pm$0.14 & 7 \\
        C/2012 F6 Lemmon & gas & 1.30$\pm$0.32 & 8 \\
        C/2014 Q2 Lovejoy & gas & 0.28$\pm$0.08 & 8 \\
        C/2007 B3 Lulin & gas & $<$1.12 & 9 \\
        153P/Ikeya-Zhang & gas & $<$0.56$\pm$0.06 & 10 \\
        \hline
        \multicolumn{4}{c}{Jupiter Family Comets} \\
        \hline
        103P/Hartley 2 & gas & 0.32$\pm$0.05 & 11 \\
        67P/Churyumov-Gerasimenko & gas & 1.06$\pm$0.14 & 12 \\
        46P/Wirtanen & gas & 0.32$\pm$0.13 & 13 \\
        45P/Honda-Mrkos-Pajdu\v{s}\'{a}kov\'{a} & gas & $<$0.40 & 14 \\
        \hline
\end{tabular}

\tablecomments{Refs: 
1. \cite{alexander2012provenances}
2. \cite{brown2012experimental}
3. \cite{bockelee1998deuterated}
4. \cite{meier1998determination}
5. \cite{villanueva2008sensitive}
6. \cite{bockelee2012herschel}
7. \cite{hutsemekers2008oh}
8. \cite{biver2016isotopic}
9. \cite{gibb2012chemical}
10. \cite{biver2006radio}
11. \cite{hartogh2011ocean}
12. \cite{altwegg201567p}
13. \cite{lis2019terrestrial}
14. \cite{lis2013herschel}
}
\label{tab:hdo_h2o_lit_ss}
\end{center}
\end{table}

\begin{table}[ht!]
\caption{Overview of HDO/H$_{2}$O values of various protostars.}
\begin{center}
\begin{tabular}{c c c c}
\hline
    \hline
        Source & Phase & HDO/H$_{2}$O & Ref. \\
        & & 10$^{-3}$ & \\
        \hline
        \multicolumn{4}{c}{Clustered Class 0 LYSOs} \\
        \hline
        NGC 1333 IRAS 4A-NW & gas & 0.54$\pm$0.15 & 1, 2 \\
        NGC 1333 IRAS 2A & gas & 0.74$\pm$0.21 & 1 \\
        NGC 1333 IRAS 4B & gas & 0.59$\pm$0.26 & 1 \\
        IRAS 16293-2422 & gas & 0.92$\pm$0.26 & 1 \\
        NGC 1333 SVS 13 & ice & $\leq$17 & 3 \\
        \hline
        \multicolumn{4}{c}{Clustered Class I LYSOs} \\
        \hline
        NGC 1333 SVS 12 & ice & $\leq$5 & 3 \\
        \hline
        \multicolumn{4}{c}{Isolated Class 0 LYSOs} \\
        \hline
        BHR 71-IRS1 & gas & 1.8$\pm$0.4 & 2 \\
        B335 & gas & 6.3$\pm$1.5 & 4 \\
        L483 & gas & 4.0$\pm$0.5 & 4 \\
        \hline
        \multicolumn{4}{c}{Isolated Class 0/I LYSOs} \\
        \hline
        L1527 IRS & ice & 4.4$^{+3.7}_{-1.7}$ & 5 \\
        \hline
        \multicolumn{4}{c}{Isolated Class I LYSOs} \\
        \hline
        V883 Ori & gas & 2.3$\pm$0.6 & 6 \\
        L1551 IRS5 & gas & 2.1$\pm$0.8 & 7 \\
        L1489 IRS & ice & $\leq$8 & 3 \\
        TMR1 & ice & $\leq$11 & 3 \\
        \hline
        \multicolumn{4}{c}{IMYSOs} \\
        \hline
        IRAS 05390-0728 & ice & $\leq$10 & 8 \\
        IRAS 08448-4343 & ice & $\leq$10 & 8 \\
        HOPS 370 & ice & 4.6$\pm$2.2 & 9 \\
        \hline
        \multicolumn{4}{c}{MYSOs} \\
        \hline
        W3 IRS5 & gas & 1.3 & 10 \\
        W33A & gas & 3.0 & 10 \\
        AFGL 2591 & gas & 3.3 & 10 \\
        NGC 7538 IRS1 & gas & 3.8 & 10 \\
        Orion KL Hot Core & gas & 3.0$^{+3.1}_{-1.7}$ & 11 \\
        NGC 6334 I & gas & 0.21$\pm$0.1 & 12 \\
        G34.26+0.15 & gas & 0.35-0.75 & 13 \\
        NGC 7538 IRS9 & ice & $\leq$8.1-11.4 & 8 \\
        GL 2136 & ice & $\leq$4 & 8 \\
        IRAS 20126 & ice & 2.6$\pm$1.4 & 9 \\
    \hline
\end{tabular}

\tablecomments{Refs: 
1. \cite{persson2014deuterium}
2. \cite{jensen2019alma}
3. \cite{parise2003search}
4. \cite{jensen2021alma}
5. this work
6. \cite{tobin2023deuterium}
7. \cite{andreu2023high}
8. \cite{dartois2003revisiting}
9. \cite{slavicinska2024jwst}
10. \cite{van2006water} (envelope-averaged abundances from radiative transfer models)
11. \cite{neill2013abundance}
12. \cite{emprechtinger2013abundance}
13. \cite{coutens2014water}
}

\label{tab:hdo_h2o_lit_ps}
\end{center}
\end{table}

\clearpage


\bibliography{paper}{}
\bibliographystyle{aasjournal}

\end{document}